\documentclass[aps,pre,amsmath,amsfonts,twocolumn,superscriptaddress]{revtex4-1}
\usepackage{ifpdf}

\usepackage{graphicx}
\usepackage{dcolumn}
\usepackage{bm,upgreek}
\usepackage{url}
\usepackage{nicefrac}

\usepackage{feynmp}
\usepackage{float}
\usepackage{microtype}
\renewcommand{\Re}{\operatorname{Re}}
\renewcommand{\Im}{\operatorname{Im}}
\newcommand{\atan}{\operatorname{atan}}

\ifpdf
\fi

\begin{document}


\title{First-order patterning transitions on a sphere as a
route to cell morphology}



\author{Maxim O. Lavrentovich}
\author{Eric M. Horsley}
\author{Asja Radja}
\author{Alison M. Sweeney}
\author{Randall D. Kamien}
\affiliation{Department of Physics and Astronomy, University of Pennsylvania, Philadelphia PA 19104, United States}


\date{\today}

\begin{abstract}
 We propose a general theory for surface patterning in many different
biological systems, including mite and insect cuticles, pollen grains,
fungal spores, and insect eggs. The patterns of interest are often
intricate and diverse, yet an individual pattern is robustly reproducible
by a single species and a similar set of developmental stages
produces a variety of patterns. We argue that the pattern diversity
and reproducibility may be explained by interpreting the pattern
development as a first-order phase transition to a spatially modulated
phase. Brazovskii showed that for such transitions on a flat,
infinite sheet, the patterns are uniform striped or hexagonal. Biological
objects, however, have finite extent and offer different
topologies, such as the spherical surfaces of pollen grains. We
consider Brazovskii transitions on spheres and show that the patterns
have a richer phenomenology than simple stripes or hexagons. We
calculate the free energy difference between the unpatterned state
and the many possible patterned phases, taking into account
fluctuations and the system's finite size. The proliferation of variety
on a sphere may be understood as a consequence of topology, which
forces defects into perfectly ordered phases. The defects are then
accommodated in different ways. We also argue that the first-order
character of the transition is responsible for the reproducibility and
robustness of the pattern formation.  
\end{abstract}


\maketitle

 \begin{figure}
\centerline{\includegraphics[height=2.2in]{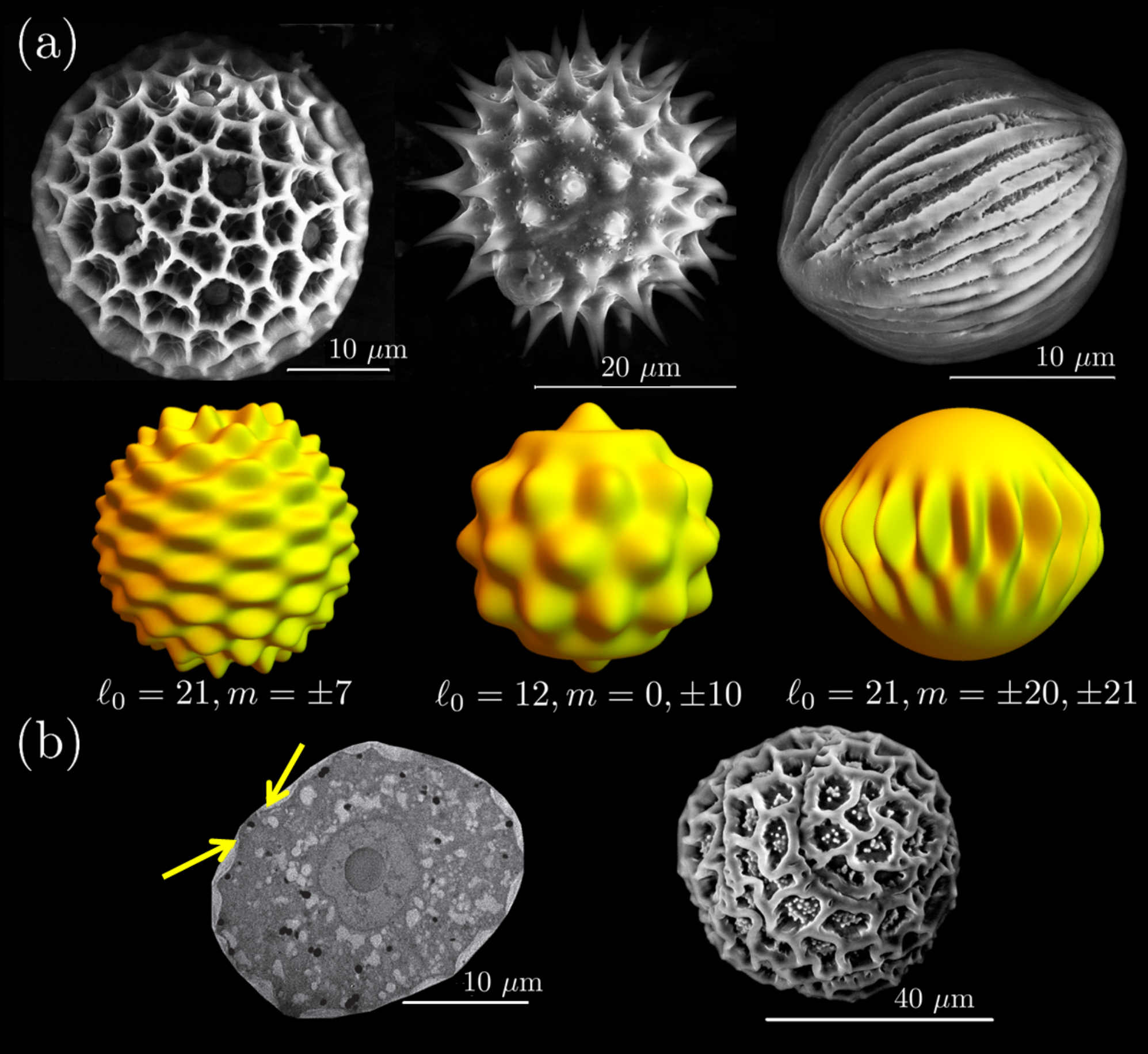}}
\caption{\label{fig:pollenpics}  (a) Electron micrographs of pollen grains.  The surface coat of the pollen, called exine, exhibits different patterns, ranging from stripes and many different patchy arrangements.  Appearing below each micrograph is a corresponding height function representation constructed from our theory with the indicated spherical harmonics. (b) The left panel shows a transmission electron microscopy cross-section of an early pollen developmental stage. The surface of the immature cell undulates (yellow arrows) with a length scale consistent with the final patterning of the mature grain shown in a scanning electron microscopy image  in the right panel.   }
\end{figure}

Surface patterning in many animal and plant species, including insect eggshells, pollen grains, fungal spores, and mite carapaces, may be extremely  diverse. Stripes, spikes, pores, ridges, and other decorations \cite{insectchapter,mitechapter}, illustrated for pollen in Fig.~\ref{fig:pollenpics}(a), all present very different geometries.  Paradoxically, the distinct morphologies may develop via the same  sequence of developmental stages \cite{pollendev1,mite1, eggshell1,palynologybook}, though  the patterns are distinctive enough to be used for taxonomic classification over eons.         In this paper, we propose a general model of the formation of these patterns, and speculate that the origin of some of these counter-intuitive features relies upon fluctuation effects leading to global pattern nucleation.

  We focus on  a class of biological surface patterns observed in many disparate taxa (fungi, arachnids, insects, angiosperms) consisting of spikes, hexagons and stripes of cross-linked polysaccharide material  tiled on a spherical cell. The surface pattern formation of these biological systems typically involves many cell components, including the cytoskeleton, plasma membrane, and cell wall (callose wall in pollen, cuticle in arthropod cuticles and fungal spores)  \cite{exinedev1, exinedev2,insectcuticlerev}.  Without some physical coupling, coordination among these many parts would require complex biological signalling across large regions of the  organism. Hence, the patterns seem more plausibly to develop via a simple physical process.  We are already familiar with complex, self-organized patterning via relatively simple processes in the natural world: convection cells at a Rayleigh-B\'enard instability \cite{HohenbergSwiftNucleation}, the patterning of pigments in animals  \cite{Turingfish}, and hexagonal patterning of dried mud or the basalt columns of the Giant's Causeway \cite{Morriscolumns}.

Whereas patterning on flat, planar substrates is expected to yield striped or hexagonal patterns \cite{Brazovskii}, we  demonstrate that the analogous transition on a sphere has a much richer phenomenology.  The spherical geometry introduces topological defects, yielding a varied set of pattern possibilities.      Also, because the transition we describe here has a first-order character, it is possible to produce a particular pattern by templating a small patch, which would then induce pattern growth over the entire surface via nucleation dynamics \cite{HohenbergSwiftNucleation}. The patterning inside the nucleation region itself could be controlled by local surface chemistry of the plasma membrane, allowing for pattern reproducibility within a species.

Although our theory may be applicable in any of the biological cases stated above, for simplicity we consider the biochemical details of pollen below, as shown in Fig~\ref{fig:pollenpics}(a), and will refer to the general case of such decorated cells as pollen. One of the earliest indications of patterning in pollen begins with plasma membrane undulations \cite{exinedev2},  as shown in Fig.~\ref{fig:pollenpics}(b). This distortion of the local membrane curvature is also implicated in other iterations of this pattern forming process, such as insect and arachnid cuticle development \cite{insectcuticlerev, butterfly}. Here, we present a  model for pattern formation via a phase transition at the plasma membrane. We show that the characteristic size of the membrane undulations, $\lambda$, is a function of physical parameters of the membrane. Hence, the membrane tension and elasticity, lipid and protein density, or osmolarity of the surrounding fluid could all vary among species and contribute to diversity in the final, observed cuticle and cell wall patterns. 

Mechanical buckling is another microscopic mechanism that may plausibly cause surface patterns in the biological systems. However, we believe our model of pattern formation may be especially applicable to  systems like pollen, since the transition to patterning may occur locally, without the homogeneous long-range forces in existing models of elastic buckling \cite{buckling1}.  Another characteristic suggestive of a phase transition is that all these systems have a cross-linked polymeric layer secreted on the surface of the cell membrane.    

We will  derive from the microscopic model a more general, coarse-grained description, which turns out to be the spherical analog of the Brazovskii model \cite{Brazovskii}. Such models describe a wide variety of systems \cite{ModPhasesOverview}, including block copolymer assembly \cite{brazovskiiExp},  crystallizing Bose-Einstein condensates in optical cavities \cite{Sarang},  and cholesteric liquid crystals \cite{BrazovskiiLC}. Such systems on a sphere might also be excellent experimental test-beds for our theory.  Although there have been recent numerical investigations of such models on a sphere via numerical methods \cite{spherediblocks2}, our analysis goes beyond this theory by incorporating fluctuations  and  provides a broader  understanding of such transitions through analytical methods. The fluctuations lead to first order behavior, suggesting a nucleation and growth scenario \cite{brazovskii2, HohenbergSwiftNucleation}.

\section{A Microscopic Model}

  As a microscopic model,  consider a concentration field $\Psi$ on the plasma membrane that might describe, for example, the concentration of a compound (or a deviation above or below some baseline value) that eventually coordinates the deposition of the tough sporopollenin exterior, e.g., the underlying primexine matrix \cite{exinedev1}.      
The pattern formation will be driven by phase separation of the concentration $\Psi$ at the plasma membrane surface.  Hence, we have a general Landau-Ginzburg  free energy  for $\Psi$:
\begin{align}
\mathcal{H}_{\Psi} & = \int \mathrm{d}^2 \mathbf{x}\,\left\{\frac{K_0}{2} | \nabla \Psi|^2+\frac{\tau_0}{2} \Psi^2+\frac{\lambda_3}{3!} \Psi^3  + \frac{\lambda_4 }{4!} \Psi^4  \right\}    \label{eq:LGden},
\end{align}
where $K_0$, and $\lambda_{3,4}$ are coupling constants that depend on the specific compound and associated biochemistry and we assume that $K_0,\lambda_4>0$.  The temperature-like parameter $\tau_0$  is quenched from positive to negative values (or below some critical value) during pattern formation.   Because the field $\Psi$  lives on a spherical surface, we use spherical coordinates $\Psi = \Psi(\theta,\phi)$ [where $\theta \in [0 ,\pi]$ and $\phi \in [0,2 \pi)$ are, respectively, the colatitude and longtiude].  The integration $\int \mathrm{d}^2 \mathbf{x}$  in Eq.~\ref{eq:LGden} is the appropriate spherical measure $\int \mathrm{d}^2 \mathbf{x} = R^2 \int \mathrm{d}\theta\mathrm{d}\phi\sin\theta$ where $R$ is the radius of the sphere.  We expand $\Psi(\theta,\phi)$,
\begin{equation}
\Psi(\theta,\phi) = \sum_{\ell=0}^{\infty} \sum_{m=-\ell}^{\ell} \Psi_{\ell}^m Y_{\ell}^m(\theta,\phi) \equiv \sum_{\bm{\ell}} \Psi_{\ell}^m Y_{\ell}^m, \label{eq:modeexpansion}
\end{equation}
where $Y_{\ell}^m\equiv Y_{\ell}^m(\theta,\phi)$ are the spherical harmonics, and  $\bm{\ell}=(\ell,m)$ is a convenient notation for their indices.  Because the scalar field $\Psi$ is real, the expansion coefficients satisfy the property $[\Psi_{\ell}^m]^*=(-1)^m \Psi_{\ell}^{-m}$.
The Landau-Ginzburg theory in Eq.~\ref{eq:LGden}  favors modes with $\ell=0$, which correspond to uniform states.  A patterned phase would prefer to have some  $\ell \neq 0$ that minimizes the free energy.  The key ingredient will be the coupling of the field $\Psi$ to the membrane curvature. The flat, infinite membrane analog of our model is studied in detail in  \cite{curvatureinstability}, which we will follow closely for our spherical model.

The membrane itself fluctuates away from its spherical shape, so that the radius varies with $\theta$ and $\phi$,  $r(\theta,\phi)=R[1+u(\theta,\phi)] $. The fluctuation field $u$ may also then be expanded in spherical harmonics with modes $u_{\ell}^m$, as in Eq.~\ref{eq:modeexpansion}.    Although there are many possible models for spherical lipid membranes, outlined in \cite{seifert}, for example, the specific form does not matter for our purposes, since the result will be general. All models will typically have a bending term with a bending rigidity $\kappa$ and a surface tension $\sigma$.  Generically, the field $\Psi$ couples to the field $u$ by introducing a \textit{spontaneous curvature}: it is reasonable that the inhomogeneity introduced by a local excess of $\Psi$ causes the membrane  to bulge in or out locally.  Apart from an irrelevant additive constant, a particular bending energy and the membrane coupling term look like
\begin{align}
\mathcal{H}_{\mathrm{mem}} & = \frac{1}{2} \sum_{\ell \geq 2,m} \Big\{ |u_{\ell}^{m}|^2(\ell+2)(\ell-1) [\kappa\ell(\ell+1)+R^2\sigma] \nonumber\\ 
& \qquad  \qquad {} - 2\mu R \ell(\ell+1)u_{\ell}^m(\Psi_{\ell}^m)^* \Big\}, \label{eq:elasticmodel}
\end{align}
where the $\ell=0$ mode is removed by constraining the total volume of the vesicle and the $\ell=1$ mode is removed because it corresponds to translations of the entire membrane. The coupling $\mu$ will depend on the microscopic details of how the spontaneous curvature is induced by the inhomogeneity.

Our total, microscopic free energy is $\mathcal{H}_{\mathrm{tot}}=\mathcal{H}_{\mathrm{mem}}+\mathcal{H}_{\Psi}$.  We can calculate thermal averages of interest using the standard Boltzmann weights.  Moreover, we can generate an effective free energy for the density field $\Psi$ by integrating out the membrane degrees of freedom.  Fortunately, because those degrees of freedom  appear at most quadratically in $\mathcal{H}_{\mathrm{tot}}$, we can perform this integration exactly, leaving an effective free energy $\widetilde{\mathcal{H}}$ for just the field $\Psi$:

 \begin{align}
  \mathcal{\widetilde{\mathcal{H}}} = \frac{1}{2} \sum_{\bm{\ell}}  \left[ \omega(\ell)+R^2\tau_0 \right] | \Psi^m_{\ell}|^2+ \mathcal{H}_{\mathrm{int}}\label{eq:quadraticH},
 \end{align}
 where $\omega(\ell)$  is now a  function of the mode number $\ell$ and the $\lambda_{3,4}$ coupling terms $\mathcal{H}_{\mathrm{int}}$ are inherited from  Eq.~\ref{eq:LGden}.  Note that for $\ell \gg 1$, 
$
 \omega(\ell) \approx \ell^2 \left[K_0- \mu^2 R^2/(\kappa \ell^2+R^2 \sigma) \right]$.

  Crucially, $\omega(\ell)$ develops a \textit{minimum} at a non-zero value of $\ell$ whenever the spontaneous curvature term is strong enough: $\mu > \sqrt{K_0 \sigma}$.  Thus, this simple coupling to membrane fluctuations leads to a spatially modulated phase with a characteristic mode number $\ell = \ell_0 \approx R[(\mu \sqrt{\sigma/K_0}-\sigma)/\kappa]^{1/2}$.  The number $\ell_0 =0,1,2,\ldots$  approximately describes the number of pattern oscillations/wavelengths that fit in a sphere circumference.  As we can see from Fig.~\ref{fig:pollenpics}, we will typically have $\ell_0 \gg 1$.  We may also relate $\ell_0$ to the characteristic wavelength $\lambda$ of the pattern, since $\ell_0 \approx 2 \pi R/\lambda$.  A rough estimate of $\lambda$ using typical parameters for lipid membranes gives  the right order of magnitude for pollen pattern features ($\lambda \sim 0.1-1$ $\upmu$m) \cite{curvatureinstability, criticalmembrane3}.

The preceding discussion shows that the effective free energy for the field modes $\Psi_{\ell}^m$ near $\ell \approx \ell_0$ has the general form
\begin{equation}
\mathcal{H}  =\frac{1}{2} \sum_{\bm{\ell}}  \left[K(\ell-\ell_0)^2+R^2\tau \right] | \Psi^m_{\ell}|^2+ \mathcal{H}_{\mathrm{int}}, \label{eq:sphBrazo}
\end{equation}
where $K$ and $\tau$ are new coupling constants that depend on the microscopic parameters in Eq.~\ref{eq:LGden} and Eq.~\ref{eq:elasticmodel} \cite{curvatureinstability}.       The interaction terms  $\mathcal{H}_{\mathrm{int}}$ continue to be inherited from Eq.~\ref{eq:LGden}. The key feature of the effective free  energy in Eq.~\ref{eq:sphBrazo} is the gradient term (the term depending explicitly on $\ell$) that is minimized when $\Psi$ is modulated on the lengthscale  $\lambda \approx 2 \pi R/\ell_0$.  This means that the physics of the pattern formation will be dominated by fluctuations at a non-zero  momentum.

Before continuing, we note that the precise microscopic model for pollen is not known, and there are many possibilities \cite{ScottReview,exinedev1}.  However, our  final result in Eq.~\ref{eq:sphBrazo} is not contingent on the particular details of our phase separation model, and we expect that the coarse-grained features of many microscopic models will obey Eq.~\ref{eq:sphBrazo}, but with different dependencies of the coupling constants $K$, $\tau$, and $\lambda_{3,4}$ on the microscopic parameters.  In any case,  the field $\Psi$ will describe the pattern template on which the tough sporopollenin material is deposited. Hence, a height function representation of this field away from a reference sphere configuration may qualitatively describe the final deposited pattern, as shown in Fig.~\ref{fig:pollenpics}(a). We will now use the final result in\ Eq.~\ref{eq:sphBrazo} to demonstrate  that robustness and variability are \textit{general} features of the pattern formation.  In the following, we set $K=1$ without loss of generality.  We begin by showing that the model generically has a first-order transition, as in the flat case \cite{Brazovskii}.

As in the flat case \cite{Brazovskii}, fluctuations will induce phase transitions to ordered states.  In preparation, we expect ordered states of the form\begin{equation}
\bar{\Psi}(\theta,\phi)= a c_0 Y_{\ell_0}^0+ \sum_{m>0} a [ c_m Y_{\ell_0}^m+c_m^*(-1)^mY_{\ell_0}^{-m}], \label{eq:realordered}
\end{equation}
where $a \geq 0$ is an overall amplitude and $c_m$ are  (generally complex) functions of $m$ that indicate the direction of the ordered state in the $2\ell_0+1$-dimensional space of $m$'s. An ordered state consisting of a single spherical harmonic mode contribution ($c_m \neq 0$ for a single $m$)  is the analog of the striped phase $\cos (k_0 \,\hat{\mathbf{k}} \cdot \mathbf{x})$  considered by Brazovskii.   The spherical harmonics encode the non-trivial topological features of the sphere.  For example,  any kind of striped ordering on a sphere must have defects according to the Poincar\'e-Brouwer theorem \cite{KamienPrimer}.
 The spherical harmonics naturally include these defects.  For example, the $m=0$ harmonics have latitudinal stripes with $+1$ defects at the poles.   Although some progress has been made in identifying what spatially modulated patterns can form on a sphere at some fixed $\ell_0$, those analyses have been largely limited to looking at particular lower order modes $\ell_0 \lesssim 12$ \cite{spherebifurcations1,spherediblocks}. We  consider the problem for general $\ell_0$.  The sphere radius $R$ will introduce a new lengthscale into the problem and finite size effects at small $R$.   In the following we  construct finite-size crossover scaling functions which capture both the large and small $R$ behavior at a fixed pattern wavelength.

\begin{figure}[t]
\centerline{\includegraphics[width=3.10in]{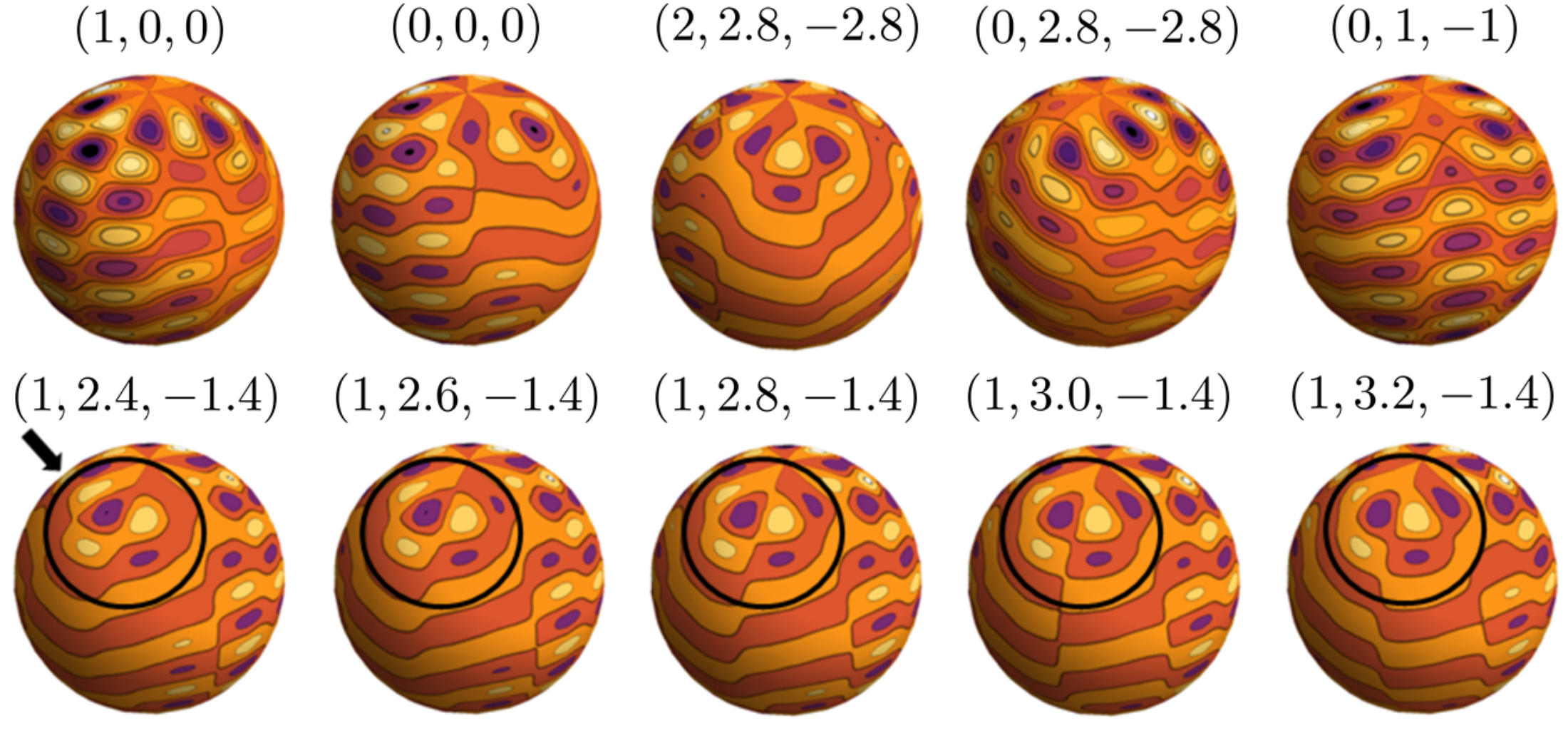}}
\caption{\label{fig:l15phases}  Plots of  metastable, ordered states (Eq.~\ref{eq:realordered}) with $\ell_0 = 15$ with identical energies on a sphere, chosen by changing the phases $e^{i \theta_m \pi/4}$ of the directions $c_m$ of the spherical harmonic modes.  The bright yellow and dark purple regions indicate, respectively, regions of maximal and minimal values of the ordered state $\bar{\Psi}$.  For these plots, we have chosen three nonzero $c_m$'s with $m=4,5,7$ and phases determined by the triple $(\theta_4,\theta_5,\theta_7)$ shown above each plot. In the bottom row, we highlight a particular region of the pattern as we vary one of the phases. Note that even though these states all have the same value of $\ell_0$ and choice of $m$'s, changing the relative phases can substantially alter the resulting pattern.}
\end{figure} 

\section{Fluctuation-Induced First Order Transition}

Consider the transition to an ordered state in our general free energy in Eq.~\ref{eq:sphBrazo}. The interaction terms $\mathcal{H}_{\mathrm{int}}$ include both a cubic and a quartic term.  A cubic term alone would induce a first-order transition to an ordered phase, which would likely be mediated via a nucleation process.  However, when $\lambda_3=0$ (see Eq.~\ref{eq:LGden}), we expect a second-order transition.  This $\lambda_3=0$ case may be especially important for our systems because it is known that the plasma membrane may tune itself to a special critical point which does not have a cubic term \cite{criticalmembrane2}. If we set $\lambda_3=0$ and pick some $\lambda_4 >0$, mean-field theory predicts a change in the character of the potential energy, $\tau \Psi^2/2+ \lambda_4 \Psi^4/4!$, when $\tau$ changes signs.  When $\tau > 0$, the potential has a minimum at $\Psi=0$.  However, when $\tau <0$, the minimum shifts to a non-zero $\Psi \propto \sqrt{-6 \tau/\lambda_4} $.  This is where we expect the ordered state to appear.     Such a transition is second-order in nature because the amplitude of the field changes continuously as we vary $\tau$.   In this situation, the patterned and un-patterned state minima never coexist and the pattern would have to develop homogeneously over the entire sphere surface, with no nucleation process.   However, we shall see that fluctuations modify this picture and instead induce a first-order transition.

To facilitate computations, it is convenient to define a  ``bare'' propagator or two-point correlation function
\begin{equation}
\left\langle \Psi_{\ell}^m \Psi_{\ell'}^{m'} \right\rangle_0= \frac{\delta_{\ell- \ell'} \delta_{m+ m'}(-1)^m}{ (\ell-\ell_0)^2+\tau R^2}\equiv\frac{\delta_{\ell- \ell'} \delta_{m+ m'}(-1)^m} {M(\ell,\tau)}  , \label{eq:barepropagator}
\end{equation}
where $\delta_x$ is the Kronecker delta function: $\delta_x=1$ if $x=0$ and $\delta_x=0$ otherwise.   The subscript $0$ on the brackets indicates that we have set the interaction terms to zero: $\mathcal{H}_{\mathrm{int}}=0$.  The terms $ \mathcal{H}_{\mathrm{int}}$ involve couplings between  different spherical harmonic modes $\Psi^m_{\ell }$, and we will have to treat these terms perturbatively.  Expanding in spherical harmonics:
\begin{align}
 \mathcal{H}_{\mathrm{int}}  
  & = \frac{ R^2}{4!}\sum_{\{\bm{\ell}_i \}_{i=1}^4} \gamma^{(4)} \prod_{j=1}^4 \Psi_{\ell_j}^{m_j} +\frac{R^2}{3!} \sum_{\{\bm{\ell}_i \}_{i=1}^3} \gamma^{(3)} \prod_{j=1}^3 \Psi_{\ell_j}^{m_j} \label{eq:quarticH}
\end{align}
with the ``bare''  vertex functions \cite{srednicki} $\gamma^{(3)} \equiv \gamma^{(3)}(\{\bm{\ell}_i \}_{i=1}^3)=\lambda_3 \Upsilon_{m_1,m_2,m_3}^{\ell_1,\ell_2,\ell_3}$, and $\gamma^{(4)} \equiv  \gamma^{(4)}( \{ \bm{\ell}_i \}_{i=1}^{4} )$ given by:
\begin{align}
\gamma^{(4)} &  ( \{ \bm{\ell}_i \}_{i=1}^{4} )=\lambda_4 \sum_{\bm{\ell}} (-1)^m\Upsilon_{m_1,m_2,m}^{\ell_1,\ell_2,\ell} \Upsilon_{m_3,m_4,-m}^{\ell_3,\ell_4,\ell}   \label{eq:bare4point0}
\end{align}
where we  have introduced a special notation for the so-called Gaunt coefficients
\begin{align}
\Upsilon_{m_1,m_2,m_3}^{\ell_1,\ell_2,\ell_3} &\equiv  \sqrt{\frac{(2 \ell_1+1)(2 \ell_2+1)(2 \ell_3+1)}{4\pi}} \nonumber \\ & \qquad \times \begin{pmatrix} \ell_1 & \ell_2 & \ell_3 \\ m_1 & m_2 & m_3 \end{pmatrix}  \begin{pmatrix} \ell_1 & \ell_2 & \ell_3 \\ 0 & 0 & 0 \end{pmatrix},
\end{align}
defined in terms of the standard Wigner $3j$-symbols \cite{stegun}, for which rapid evaluation  algorithms are available \cite{numgaunt}.  We  follow Brazovskii's calculation and make use of a Hartree-Fock (HF) approximation in which the corrections due to fluctuations are calculated self-consistently using a particular subset of Feynman diagrams.  The details of the calculation are given in the \textit{SI Text}.  We  always work in the limit that the coupling coefficients $\lambda_{3,4}$ are small.

The  HF approximation of the renormalized propagator is written as a self-consistency condition on  $t_d$, the fluctuation-renormalized value of $\tau$  in the disordered state:
\begin{equation}
 t_d - \tau= \frac{\lambda_4 }{8\pi } \sum_{\ell}{}^{'} \frac{2 \ell+1}{M(\ell,t_d)}  \approx   \frac{\lambda_4 \ell_0}{4R \sqrt{t_d} }\coth \left(\pi R  \sqrt{t_d} \right) \label{eq:propcrossover} 
\end{equation}
The summation over $\ell$ in Eq.~\ref{eq:propcrossover} is the discrete analog of an integration of the propagator over all modes (i.e.,  a one-loop correction).  The prime on the sum indicates a regularization procedure where the divergence associated with large $\ell$ is removed. The specific regularization procedure only modifies the short wavelength (large $\ell$) physics, and is irrelevant for the coarse-grained features of the pattern formation. Also, we expect that the contribution from the cubic interaction is negligible for $\ell_0 \gg 1$ (see \textit{SI text}). Note that the function in Eq.~\ref{eq:propcrossover}  captures both a large radius regime, $\pi R \sqrt{ t_d} \gg 1$  and a small radius regime $\pi R \sqrt{t_d} \ll 1$. Thus, the correction crosses over to a finite-size dominated behavior when the  correlation length $\xi \approx 1/\sqrt{t_d}$ of fluctuations becomes  large compared to the sphere's pole-to-pole distance: $\xi \gg \pi R$.

Equation~\ref{eq:propcrossover} admits only positive solutions for $t_d$ for any value of $\tau$.  Hence, fluctuations prevent the temperature-like term from changing sign.  If a cubic term were present, then a first-order transition is possible if $t_d$ is sufficiently small.  However, if $\lambda_3=0$, then the only possibility for \textit{any} transition is if the quartic term proportional to $\lambda_4$ is driven negative.      We must therefore consider the $\lambda_3=0$ case in more detail to find which modes have a fluctuation-induced sign change in the quartic term. 

Turning to the 4-point vertex function $\Gamma^{(4)}(\bm{\ell}_1,\bm{\ell}_2,\bm{\ell}_3,\bm{\ell}_4)$, we can see that the modes of interest with the largest fluctuation effects all have  $\ell=\ell_0$, as readily seen in the propagator expression in Eq.~\ref{eq:barepropagator} where the denominator is smallest near $\ell = \ell_0$.  Thus, we focus on the particular  vertex function $\Gamma^{(4)}_{\ell_0} \equiv \Gamma^{(4)}(m_1,-m_1,m_2,-m_2)$, corresponding to the coupling constant of quartic terms of the form   $|\Psi^{m_1}_{\ell_0}|^2 | \Psi^{m_2}_{\ell_0}|^2$.
 In the one-loop HF approximation, in the absence of a cubic term $(\lambda_3=0),$ the vertex function $\Gamma^{(4)}_{\ell_0}$ is given by\begin{align}
\Gamma^{(4)}_{\ell_0}& =  \sum_{\ell} \frac{\lambda_4}{1+\lambda_4 \Pi(\ell)}  \bigg[   \Upsilon_{m_1,- m_1,0}^{\ell_0,\ell_0,\ell} \Upsilon_{m_2, -m_2,0}^{\ell_0,\ell_0,\ell}-\lambda_4\Pi (\ell)  \nonumber \\ &  {}  \times(\Upsilon_{m_1,m_2, 0}^{\ell_0,\ell_0,\ell} \Upsilon_{-m_1, -m_2,0}^{\ell_0,\ell_0,\ell} +\Upsilon_{m_1,-m_2, 0}^{\ell_0,\ell_0,\ell} \Upsilon_{-m_1, m_2,0}^{\ell_0,\ell_0,\ell}) \bigg] , \label{eq:renorm4point}
\end{align}
where  $\Pi(\ell)>0$ is an integration over a product of \textit{two}  propagators:
\begin{equation}\Pi(\ell) \approx \frac{R^2\Upsilon_{0,0,0}^{\ell_0,\ell_0,\ell}}{4\sqrt{\pi(2\ell+1)}} \sum_{\ell_{1,2}} \prod_{i=1}^2\sqrt{2\ell_i+1}\, M^{-1}(\ell_i,t_d) \label{eq:vertexloop}
\end{equation}

The three $m$-dependent Gaunt coefficient terms in Eq.~\ref{eq:renorm4point} are three different angular momentum ``channels'' which  contribute to the vertex. A single momentum channel contributes whenever $m_1 \neq \pm m_2$, so that the two terms in the second line of Eq.~\ref{eq:renorm4point} vanish.  Then, the renormalized vertex $\Gamma^{(4)}_{\ell_0}$ has the same sign as the bare vertex $\gamma^{(4)}$ in Eq.~\ref{eq:bare4point0} (since $\lambda_4 \Pi(\ell)>0$ for all $\ell$).   However, if $m_1=\pm m_2$,  then one of the other two channels start to contribute.  There is also a special case for which all three channels contribute: $m_1=m_2=m_3=m_4=0$.  Note from the second line of Eq.~\ref{eq:renorm4point} that if two or more channels contribute and if $\lambda_4 \Pi(\ell)>1$, the renormalized vertex function changes sign!  This indicates the possibility of a first order transition for these $m_1=\pm m_2$ modes with  $\ell=\ell_0$.  They are, in fact, the modes we have considered already in Eq.~\ref{eq:realordered} and are the spherical analogs of the cosine standing waves of the flat space Brazovskii analysis.

We now examine the most divergent piece of the fluctuation correction $\Pi(\ell)$ to see if we generically expect that $\lambda_4 \Pi(\ell)>1$.  The most divergent part of the correction occurs when $\ell_1=\ell_2 \approx \ell_0$ in Eq.~\ref{eq:vertexloop}. Setting $\ell_1=\ell_2$, we find that $\Pi(\ell)$ diverges as $t_d \rightarrow 0$ as $\Pi(\ell) \sim t_d^{-3/2}$ in the  planar limit ($\pi R \sqrt{t_d} \gg 1$ with $\ell_0/R=k_0$ fixed) and as  $\Pi(\ell) \sim t_d^{-2}$   in the finite size limit $\pi R \sqrt{t_d} \ll 1$. Thus, because $\Pi(\ell) \rightarrow \infty$ as $t_d \rightarrow 0$,  the vertex function for the special modes in Eq.~\ref{eq:realordered} is expected to change sign due to fluctuations, consistent with the Brazovskii result.   

We have now shown that our model generically exhibits a first-order transition to a patterned phase.  In the absence of a cubic term in the  terms $\mathcal{H}_{\mathrm{int}}$, this transition is particularly interesting as the first-order character is induced by fluctuations.  We  now calculate the free energies of the ordered states.  We will find that differences between plane waves in the plane and spherical harmonics on the sphere lead to a much richer variety of possible states -- the ``zoo'' of pollen patterns!

\section{Patterned States}

We now consider an ordered  state $\bar{\Psi}$ that minimizes the thermodynamic potential with nonvanishing spherical harmonic coefficients $\bar{\Psi}_{\ell}^m$. We expand our field around this state,  $\Psi_{\ell}^m = \psi_{\ell}^m + \bar{\Psi}_{\ell}^m$,
where $\psi_{\ell}^m$  are the fluctuations around the potential minimum $\bar{\Psi}$, i.e.,  $\langle \psi_{\ell}^m \rangle=0$.  To determine whether an ordered state is more stable than a disordered state, we need to generate the effective free energy as a function of the average field configuration, $W[\bar{\Psi}]$.  To do this, we add an external field $h$ to $\mathcal{H}$, and calculate the partition function as a function of $h$ to generate the free energy, $F[h]$.  A Legendre transform $W[\bar{\Psi}] = F[h] + \int \mathrm{d}^2x\,h\bar{\Psi}$, where $h$ satisfies $\bar{\Psi} = - \delta F/\delta h$, generates $W[\bar{\Psi}]$ -- from this we can calculate the free energy of various states $\bar{\Psi}$.   This is difficult to implement, so we follow Brazovskii's ingenious approximation method for calculating the free energy difference per unit area, $\Delta \Phi$, between the ordered and disordered states.

Through a change of variables in the functional integral for the partition function, we expand $\mathcal{H}$ in powers of $\psi$ around $\bar{\Psi}$ resulting in a theory for the modes $\psi_{\ell}^m$, the fluctuating degrees of freedom. We then relate $h$ to $\bar\Psi$ to lowest order, leading to a linearized theory for $h$ \cite{Brazovskii}.  Because the unstable modes have $\ell = \ell_0$, we may parameterize the modes as in Eq.~\ref{eq:realordered}: $\bar{\Psi}_{\ell_0}^m\equiv\bar{\Psi}_m = a c_m$.   We are now set to calculate the free energy change $\Delta \Phi$ between the disordered and patterned states.   
To do this, we start in the disordered state where $\bar{\Psi}=0$ and apply an external field $h$ to tilt the potential so that, for $h$ large enough, the ordered state becomes the minimum, and then return $h$ to $0$.  During this process, the amplitude $a$\ changes from\ $a=0$ to a final $a = \bar{a}$. The final state must also be an extremum of the free energy at $h=0$ -- another minimum.  The difference in free energy then tells us whether the ordered state is more or less stable than the disordered state.

A field $h$ in the direction of the state $\bar{\Psi}$ will have spherical harmonic modes $h_m$ (with $\ell=\ell_0$) that couple linearly to $\bar{\Psi}_m$ in the free energy.   An equation  of state for $h_m$ is constructed   by differentiating the average free energy per unit area $\Phi$ with respect to  $\bar{\Psi}_m$.  Dropping terms using  $\langle \psi_{\ell}^m \rangle=0$, as well as terms of the form $\langle \psi_{\ell_1}^{m_1} \psi_{\ell_2}^{m_2} \psi_{\ell_3}^{m_3}\rangle$, which we expect to be small for similar reasons as in the Brazovskii analysis \cite{Brazovskii}, we have:
  \begin{equation}
h_{m}  =\frac{\delta \Phi}{\delta \bar{\Psi}_m} =\frac{1}{4\pi R^2} \Big\langle \frac{\delta \mathcal{H}[\psi+\bar{\Psi}]}{\delta \bar{\Psi}_{m}} \Big\rangle  ,   \label{eq:generalhMT} 
\end{equation}
where the average is taken with respect to the Hamiltonian without an applied field. A detailed expansion in terms of $\bar{\Psi}$ can be found in the \textit{SI Text}, Eq.~S59.
To simplify calculations and facilitate analytic solutions, we consider the states which satisfy this condition by pairwise cancellation of two of the modes, e.g., via $m_1=-m_2$ and $m_3=0$.

Calculating the fluctuation-corrected free energy requires the fluctuation-corrected two-point function $ g \equiv g(\bm{\ell}_1,\bm{\ell}_2)=\langle \psi_{\ell_1}^{m_1}  \psi_{\ell_2}^{m_2} \rangle$.   In the self-consistent HF approximation we have  
\begin{align}
&g^{-1}(\bm{\ell}_1,\bm{\ell}_2)  =M(\ell,\tau)(-1)^{m_2} \delta_{\ell_1-\ell_2} \delta_{m_1+m_2} +\frac{ R^2}{2} \sum_{ \bm{\ell}_{3,4}} \Big\{ \nonumber \\ & {}\qquad\gamma^{(4)}( \{ \bm{\ell}_i \}_{i=1}^{4} )\left[ g(\bm{\ell}_3,\bm{\ell}_4)+ \delta_{\ell_3-\ell_0}\delta_{\ell_4-\ell_0}   \bar{\Psi}_{m_3} \bar{\Psi}_{m_4} \right] \Big\}. \label{eq:orderedprop}
\end{align}
The major difference between this propagator and the disordered state propagator is the presence of the term proportional to $\bar{\Psi}_{m_3} \bar{\Psi}_{m_4}$.  This ordered state term introduces a dependence on the  directions $m_{1,2}$.  There are also off-diagonal terms with $m_1 \neq - m_2$.  These contributions may be ignored as long as $\tau$ is sufficiently small \cite{Brazovskii}, which we  assume in the following.

Substituting Eq.~\ref{eq:orderedprop} into Eq.~\ref{eq:generalhMT} and making an isotropic approximation to the propagator $g^{-1}$ \cite{anisotropicBrazo}, we eliminate the $\tau$-dependence in Eq.~\ref{eq:generalhMT}, leaving the following  equation of state:
\begin{align}
h_{m} &=\left[t   +\frac{\lambda_4(\delta_m-3)\alpha_{m,m}^{\ell_0}}{24\pi }|\bar{\Psi}_{m}|^2  \right] \frac{\bar{\Psi}_{m}^*}{4 \pi}+  \frac{\lambda_3}{4\pi} \sum_n(-1)^n  \nonumber \\
&\qquad  \times \Upsilon^{\ell_0,\ell_0,\ell_0}_{n,-n,0}\left[\bar{\Psi}_n\left(\frac{1}{2}-\delta_n\right)\delta_m+ \bar{\Psi}_0\delta_{m-n}\right]\bar{\Psi}_n^*,  \label{eq:ampequation}
\end{align}
where we define a convenient new variable $\alpha_{m_1,m_2}^{\ell_0}=4\pi\sum_{\bar{\ell}} (-1)^{m_1+m_2}\Upsilon^{\ell_0,\ell_0,\bar{\ell}}_{m_1,-m_1,0} \Upsilon^{\ell_0,\ell_0,\bar{\ell}}_{m_2,-m_2 ,0}$ and a renormalized temperature parameter $t$ that satisfies the equation
 \begin{equation}
t= \tau+\frac{ \lambda_4\ell_0}{4 R\sqrt{t}} \, \coth(\pi R  \sqrt{t})+ \frac{\lambda_4}{8 \pi} \sum_{m}  |\bar{\Psi}_{m}|^2. \label{eq:massgeneralMT}
 \end{equation}  
Note that when we are in the disordered state, $\bar{\Psi}=0$, then\ $t = t_d$, and  Eq.~\ref{eq:massgeneralMT} reduces to Eq.~\ref{eq:propcrossover}.  In the ordered state, we find a different temperature-like parameter $t = t_o$.

\begin{figure}[t]
\centerline{\includegraphics[width=3.1in]{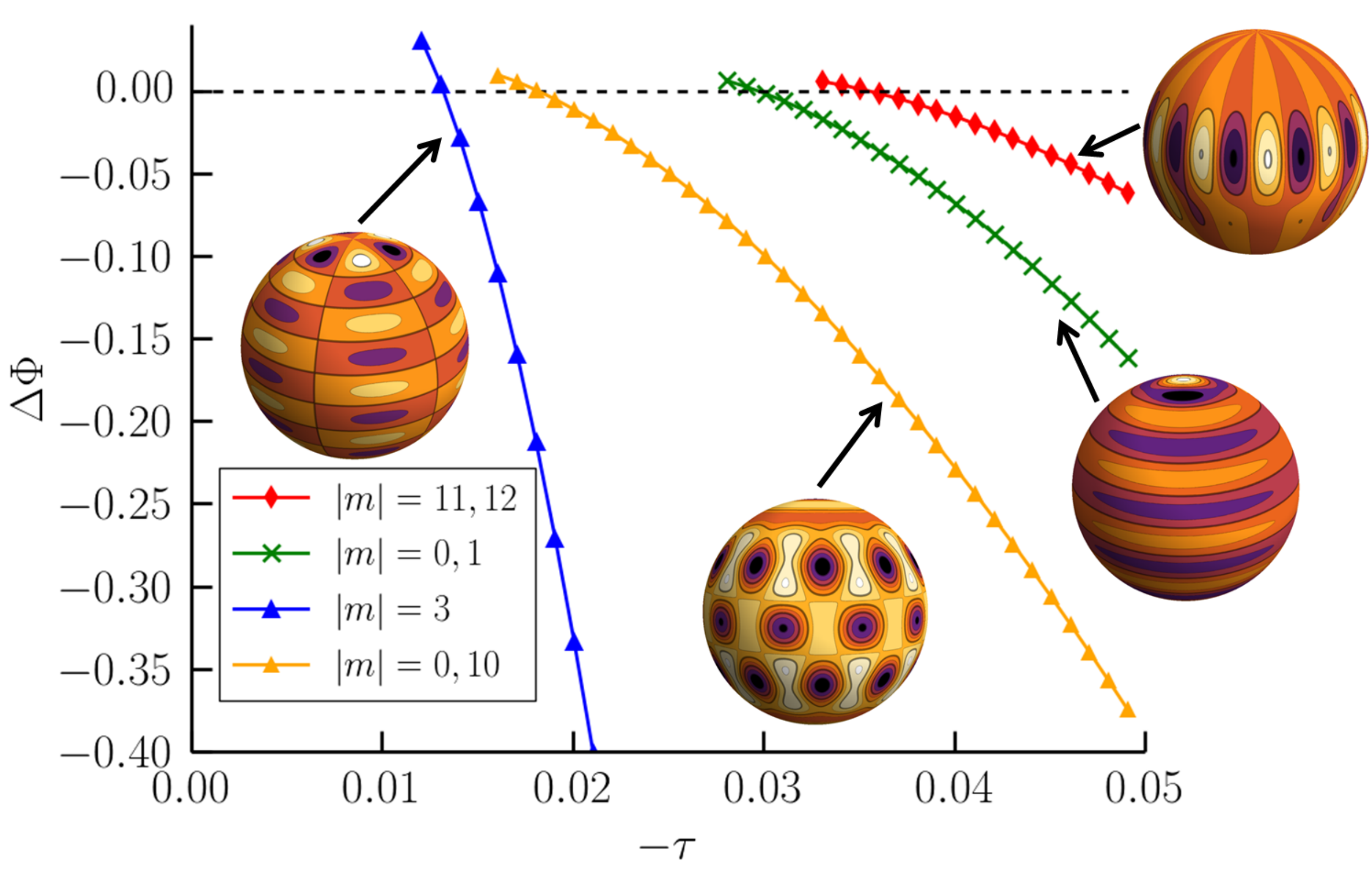}}
\caption{\label{fig:deltaphis} The free energy difference $\Delta\Phi$ between ordered states and the disordered phase as a function of the reduced temperature $\tau\ <0$ for $\ell_0=12$, $R=10$, and $\lambda_4=0.01$. The plot legend shows the chosen combination of $m$'s. The cubic term coefficient is zero except for the $|m|=0,10$ case, for which $\lambda_3=0.015$. When $\lambda_3=0$, single $|m|\approx\ell_0/2$ modes are favored for these modest values of $\ell_0$. At higher values of $\ell_0$, we find that linear combinations are more favorable, instead. The presence of a cubic term favors the formation of phases with hexagonal patterns. As we decrease the temperature (increasing $-\tau$), the ordered states become more favorable. There are a wide variety of metastable ordered states.      }
\end{figure}

Now we compute the change in free energy $\Delta \Phi$.   In the disordered state  $t=t_d$, where $t_d$ satisfies Eq.~\ref{eq:propcrossover}.  We parameterize $\bar{\Psi} = a c_m$ through an amplitude $a$ that will increase from $a=0$ to  $a = \bar{a}$.   Because the final state must correspond to a free energy minimum after the field $h$ is turned off,    $\bar{\Psi}_m = \bar{a} c_m$ must satisfy Eq.~\ref{eq:ampequation} with $h_m=0$ for all $m$.  A convenient choice for the final amplitude is $\bar{a}^2 = 4 \pi t_o/\lambda_4$.  The coefficients $c_m$  are  calculated by setting $h_m=0$ and $\bar{\Psi}_m = \bar{a} c_m $ in Eq.~\ref{eq:ampequation}.  In the absence of a cubic term ($\lambda_3=0$), the solution is particularly simple.  Either $c_m=0$ or $|c_m|^2= 6/[(3-\delta_{m})\alpha^{\ell_0}_{m,m}]$. 
Note that only the magnitude $|c_m|$ of the mode directions is specified.  Thus,  at this order of perturbation, ordered states with different relative phases in the $c_m$'s have identical energies.    Patterns on a flat, infinite, substrates have a similar degeneracy, but the  phases do not strongly modify the pattern  \cite{leibler}.  For the sphere, the relative phases  generate markedly different patterns due to the presence of defects, as shown in  Fig.~\ref{fig:l15phases}.  Corrections to our approximation ( e.g., higher order terms in Eq.~\ref{eq:quarticH}) may break the degeneracy, but many  patterns are likely nearly degenerate on a sphere.  When $\lambda_3 \neq 0$, the coefficients $c_m$ may be found numerically, but, again, we find that only the magnitudes $|c_m|$ are specified for the $m>0$ coefficients. Hence,   there remains a large degeneracy of possible patterns due to the relative phase freedom {\it even when the cubic term is included}: The presence of explicit symmetry breaking does not alter the conclusion that pattern formation on the sphere is  qualitatively different than that on the plane.

We may construct ordered states with arbitrary numbers of non-zero $c_m$'s but only those combinations with $\Delta\Phi<0$ for some negative value of $\tau$ correspond to stable patterns. Integrating up the free energy changes, we find
\begin{equation}
\Delta \Phi = \frac{1}{4\pi R^2}   \int_0^{\bar{a}} \, \left\langle\frac{\partial  \mathcal{H}}{\partial a} \right\rangle\mathrm{d}a=\sum_m \int_0^{\bar{a}} h_m \frac{\partial\bar{\Psi}_{m} }{\partial a}\mathrm{d}a, \label{eq:generaldeltaphi}
\end{equation}
where $\partial_a \bar{\Psi}_m = c_m$ for our parameterization of the ordered states.  Substituting Eq.~\ref{eq:generalhMT} into Eq.~\ref{eq:generaldeltaphi} yields a complex expression for $\Delta\Phi$ (shown in \textit{SI Text}, Eq.~S69) -- finding the values of $c_m$ for which $\Delta\Phi$ is negative allows us to find preferred ordered states.  As an example, we plot $\Delta \Phi$ for different ordered states with $\ell_0=12$  in Fig.~\ref{fig:deltaphis}. 

Roughly speaking, when $\lambda_3=0$ the most favored ordered states are ones for which $\sum_m |c_m|^2 \gtrsim 2$.  For modest $\ell_0 \sim 10$, we find that single mode solutions with $m \approx \ell_0/2$ work best, as illustrated in Fig.~\ref{fig:deltaphis}.  At higher values  $\ell_0 \gtrsim 30$, the latitudinal and longitudinal striped solutions with two modes  ($m \approx 0,1$ and $m=\ell_0-1,\ell_0$, respectively) work best.  For even larger $\ell_0$, the coefficients $\alpha_{m,m}^{\ell_0}$  behave like $\alpha_{m,m}^{\ell_0} \sim \ln \ell_0$.  This means that the  ordered states have more modes, allowing for the possibility of different patterns with (nearly) degenerate energies (see Fig.~\ref{fig:l15phases}).  In the presence of a cubic term, hexagonally-patterned states are favored, as shown for the $|m|=0,10$ case in Fig.~\ref{fig:deltaphis}. These states also have defects  and resemble those found in the absence of fluctuations \cite{spherediblocks2}. In all these cases, choosing different values for $\ell_0$ yields qualitatively different  stable patterns. This is in  contrast to the planar case, where  striped or hexagonal solutions are favored for any $k_0$.

  Because multiple modes contribute to the ordered state for large $\ell_0$ and the choice of phase for $c_m$ (Fig.~\ref{fig:l15phases}) influences the resulting pattern, we expect a rich phase structure.   Further, at large radii $R$ and fixed pattern wavelength $\lambda \approx 2\pi R/\ell_0$, the single-mode, uniform stripe solutions with two $+1$ defects at the poles are {\sl not} favored in our approximation.  One possibility is that the  ordered states are spiral-like \cite{spherediblocks2} (four $+1/2$ defects), which would require an analysis of adjacent modes  $\ell =\ell_0 \pm 1$ \cite{spirals}.  Of course, regardless of sphere size, the defects are always present and may be accommodated in different ways. As a result, determining the precise phase diagram and minimum energy states is beyond the scope of this  approach, which focuses on one value of $\ell$.  This should be contrasted with the plane, where the minimum energy ordered states are defect-free and the phase diagram can be more readily constructed.  Finally, many different ordered states yield a negative $\Delta \Phi$ (see Fig.~\ref{fig:deltaphis}),  i.e., many different patterns are  metastable.  So, pollen may, for example, locally apply a field $h$ via a biochemically-controlled process to force the pattern into a particular metastable ordered state.  The pollen may then ``quench'' this pattern, forcing it to spread over the surface via a nucleation process.

\section{Conclusions}

In conclusion, we have developed a phenomenological theory of pattern formation on a sphere. This theory provides a plausible explanation of the physical origins of micron-scale surface textures found on cell walls and cuticles of distantly related taxa such as plants, mites, fungi and insects. We showed how this mechanism may originate in plasma membrane undulations coupled to the phase separation of polysaccharide materials, which later coordinate the deposition of a tough exterior wall.  Our theory predicts that the pollen grain surface is quenched below a first-order transition point during development, and have argued that a patterned phase can spread after the quench via a nucleation process.   A given species may specify one of these many patterned modes via a  nucleation site defined by one or more of several possible cell-biological mechanisms. For example, a localized  site could be designated by the local surface chemistry of the plasma membrane relative to one pole of the cell, or by  crowding at the cell surface of nascent pollen caused by  ordered packing in the developing anther.  We showed that the first-order character of the transition will be maintained even when the free energy has no cubic term.   We also argued that the theory without a cubic term may be particularly relevant because the plasma membrane composition \textit{in vivo} may be tuned to a critical point \cite{criticalmembrane2}.

Whereas the first-order character of this transition may explain the reproducibility of a pattern in one species, the theory may also provide an answer to why there is so much pattern variability among different species.  First, a wide variety of patterns is possible by modifying the nucleation pattern, which, once formed, allows the rest of the pattern to propagate rapidly and robustly across the surface.  Second, pattern formation on a sphere is intrinsically varied because, in contrast to the planar case, the ordered states on the sphere must accommodate  defects, providing a larger space of possible patterns. By contrast, butterfly wing scale development may be an example of patterning on a flat substrate via this mechanism; the distal surface of the wing scale forms exclusively striped patterns and the plasma membrane has also been implicated in the initial pattern templating    \cite{butterfly}.

There is much room for future work:  A detailed phase diagram might be constructed using numerical techniques described in Ref.~\cite{spherediblocks2} and incorporating  fluctuation corrections. It would also be helpful to study the dynamics in order to understand how a nucleation region might be specified, leading to a particular global pattern.  There has already been progress on this in the planar case \cite{HohenbergSwiftNucleation}, providing a starting point for the spherical case.

\begin{acknowledgments}
We thank S. A. Brazovskii, S. Gopalakrishnan and D. Audus for encouraging comments and valuable discussions.
  This work was supported, in part, by the National Science Foundation through grant DMR-1262047 (R.D.K.), a Packard Foundation Fellowship to A.M.S., and a Kaufman Foundation New Research Initiative award.   R. D. K. was partially supported by a Simons
Investigator grant from the Simons Foundation.\end{acknowledgments}

\appendix

\begin{fmffile}{SIFeyn}

\section{Useful Identities and Relations}

In this Appendix we collect all the relevant Gaunt coefficient identities used in the calculations.
Recall that the Gaunt coefficients in the main text were defined as follows
\begin{align}
\Upsilon_{m_1,m_2,m_3}^{\ell_1,\ell_2,\ell_3} &  \equiv  \sqrt{\frac{(2\ell_1+1)(2\ell_2+1)(2\ell_3+1)}{4 \pi}} \nonumber \\ & \qquad \times \begin{pmatrix} \ell_1 & \ell_2 & \ell_3 \\ m_1 & m_2 & m_3 \end{pmatrix}\begin{pmatrix} \ell_1 & \ell_2 & \ell_3 \\ 0 & 0 & 0 \end{pmatrix},\label{eq:gaunt}
\end{align}
where the 2 by 3 matrices are the Wigner $3j$-symbols, which are related to the Clebsch-Gordon coefficients used for adding angular momenta in quantum mechanics \cite{stegun}. We will now derive various identities for the Gaunt coefficients from the known properties of the Wigner $3j$-symbols, which are familiar from the quantum mechanics literature. 

The Gaunt coefficients  which appear in Eq.~\ref{eq:gaunt} satisfy triangle relations given by
\begin{align}
\Upsilon_{m_1,m_2,m_3}^{\ell_1,\ell_2,\ell_3}&=0\, \, \text{if}\, \, \ell_3>|\ell_1+\ell_2| \, \, \text{or} \, \, \ell_3<|\ell_1-\ell_2| \label{trirule} \\
\Upsilon_{m_1,m_2,m_3}^{\ell_1,\ell_2,\ell_3}&=0\, \, \text{if}\, \, m_1+m_2+m_3 \neq 0 \label{msum}
\end{align}
Furthermore, because the $3j$-symbol is invariant under an even permutation of its columns, and an odd permutation generates an overall factor of $(-1)^{\ell_1+\ell_2+\ell_3}$, the presence of two such symbols in the Gaunt coefficients means that the latter coefficients are invariant under any permutation of the indices, i.e. $\Upsilon_{m_1,m_2,m_3}^{\ell_1,\ell_2,\ell_3}=\Upsilon_{m_1,m_3,m_2}^{\ell_1,\ell_3,\ell_2}=\Upsilon_{m_3,m_2,m_1}^{\ell_3,\ell_2,\ell_1}=\ldots$.
The second $3j$-symbol in Eq.~\ref{eq:gaunt} has a special from and implies the following selection rule:
\begin{equation}
\Upsilon_{m_1,m_2,m_3}^{\ell_1,\ell_2,\ell_3}=0 \, \, \text{if} \, \, \ell_1+\ell_2+\ell_3 \, \, \text{is odd}. \label{sumleven}
\end{equation}
The Gaunt coefficients also obey a reflection property (again due to a similar property of the $3j$-symbol):
\begin{equation}
\Upsilon_{m_1,m_2,m_3}^{\ell_1,\ell_2,\ell_3}=\Upsilon_{-m_1,-m_2,-m_3}^{\ell_1,\ell_2,\ell_3}.
\end{equation}
Finally, the following special case will be useful:
\begin{equation}
\Upsilon^{\ell_1,\ell_2,0}_{m_1,m_2,0} = \frac{\delta_{\ell_1-\ell_2} \delta_{m_1+m_2}(-1)^{m_1}}{\sqrt{4\pi}}. \label{eq:specialGval}
\end{equation}
We use the same convention for $\delta_x$, the Kronecker delta function, as was used in the main text: $\delta_x=1$ if $x=0$ and $\delta_x=0$ otherwise.

Like the $3j$-symbols, the Gaunt coefficients obey various summation relations.  The first one of interest is on the quantum numbers on the bottom row for one coefficient,
\begin{equation}
\sum_{m_1}  (-1)^{m_1}   \Upsilon_{m_1,-m_1,0}^{\ell_1,\ell_1,\ell}=\frac{(2\ell_1+1) \delta_{\ell}}{\sqrt{4\pi}} \label{eq:singlemsum}
\end{equation}
 and for two of them: 
 \begin{align}
 \sum_{m_{1,2}}\Upsilon_{m_1,m_2,m_3}^{\ell_1,\ell_2,\ell_3} \Upsilon_{m_1,m_2,m_3'}^{\ell_1,\ell_2,\ell_3'} &=\Upsilon_{0,0,0}^{\ell_1,\ell_2,\ell_3}\delta_{\ell_3-\ell_{3}'} \delta_{m_3-m_3'} \nonumber \\ & \times \sqrt{\frac{(2\ell_1+1)(2\ell_2+1)}{4\pi(2 \ell_3+1)}}. \label{eq:sumtwoms}
 \end{align}

To expand the cubic and quartic terms in our Hamiltonian $\mathcal{H}$ (terms proportional to $\lambda_{3,4}$ in Eq.~\ref{eq:LGden} in the main text), it is necessary to compute the integral of a product of three and four spherical harmonics $Y^{m}_{\ell} \equiv Y^m_{\ell}(\theta,\phi)$ ($\ell=0,1,2,\ldots$; $m=-\ell,-\ell+1,\ldots,\ell$) over the spherical coordinates $\theta$ (colatitude) and $\phi$ (longitude).  To make our notation more compact, we introduce a vector of indices $\bm{\ell}\equiv (\ell,m)$, so that summations over the indices may be written as follows: 
\begin{equation}
\sum_{\bm{\ell}}\equiv \sum_{\ell}\sum_{m=-\ell}^{\ell}. \nonumber  
\end{equation}
The integral of three spherical harmonics is known to be:
\begin{equation}
\int \mathrm{d}\Omega \, Y_{\bm{\ell}_1}Y_{\bm{\ell}_2} Y_{\bm{\ell}_3}=\Upsilon_{m_1,m_2,m_3}^{\ell_1,\ell_2,\ell_3}.
\end{equation}
With this one can immediately write down the expansion of the cubic term,
\begin{align}
\int \mathrm{d}\Omega \, \Psi^3 =& \sum\limits_{\{\bm{\ell}_i\}_{i=1}^3} \Psi_{\bm{\ell}_1}\Psi_{\bm{\ell}_2}\Psi_{\bm{\ell}_3} \int \mathrm{d}\Omega\, Y_{\bm{\ell}_1}Y_{\bm{\ell}_2} Y_{\bm{\ell}_3}  \\
=&\sum\limits_{\{\bm{\ell}_i\}_{i=1}^3} \Upsilon_{m_1,m_2,m_3}^{\ell_1,\ell_2,\ell_3}\Psi_{\bm{\ell}_1}\Psi_{\bm{\ell}_2}\Psi_{\bm{\ell}_3} \nonumber
\end{align}
The product of four spherical harmonics is expanded using the following identity:
\begin{equation}
\begin{split}
Y_{\ell_1}^{m_1}(\theta,\phi)Y_{\ell_2}^{m_2}(\theta,\phi)=&\sum_{\bm{\ell}} \Upsilon_{m_1,m_2,m}^{\ell_1,\ell_2,\ell} (Y_{\ell}^{m})^* \label{sphharmproduct}
\end{split}
\end{equation}
So, the quartic term reads
\begin{align}
\int \mathrm{d}\Omega \, \Psi^4 =& \sum\limits_{\{\bm{\ell}_i\}_{i=1}^4} \Psi_{\bm{\ell}_1}\Psi_{\bm{\ell}_2}\Psi_{\bm{\ell}_3}\Psi_{\bm{\ell}_4} \int \mathrm{d}\Omega\, Y_{\bm{\ell}_1}Y_{\bm{\ell}_2} Y_{\bm{\ell}_3}Y_{\bm{\ell}_4} \label{eq:vertexl1} \\
=&\sum\limits_{\{\bm{\ell}_i\}_{i=1}^4,\bm{\ell},\bm{\ell}'} \Upsilon_{m_1,m_2,m}^{\ell_1,\ell_2,\ell}\Upsilon_{m_3,m_4,m'}^{\ell_3,\ell_4,\ell'} \nonumber \\ & \qquad \qquad  \times \int \mathrm{d} \Omega \,(Y_{\ell}^{m})^*(Y_{\ell'}^{m'})^*\, \prod_{i=1}^4\Psi_{\ell_i}^{m_i} \label{eq:vertexl2}\\
=&\sum\limits_{\{\bm{\ell}_i\}_{i=1}^4,\bm{\ell}}(-1)^m\Upsilon_{m_1,m_2,m}^{\ell_1,\ell_2,\ell}\Upsilon_{m_3,m_4,-m}^{\ell_3,\ell_4,l}\prod_{i=1}^4\Psi_{\ell_i}^{m_i} .  \label{eq:quarticintegral}
\end{align}
Note that the pairing off of the spherical harmonic modes $Y_{\ell_i}^{m_i}$ modes  in Eq.~\ref{eq:vertexl1} is arbitrary.  Hence, we may rearrange the $m_i$'s ($i=1,\ldots,4$) in the two Gaunt coefficients in Eq.~\ref{eq:quarticintegral} any way we like.  This will be an important symmetry of these Gaunt coefficients which we will use when calculating the loop corrections in the next section.

\section{The Disordered State and Loop Corrections \label{appxdisordered}}
We now calculate the 2-point correlation function or propagator $g_d$ and 4-point vertex function $\Gamma^{(4)}$ in the disordered phase.  We put a subscript on the propagator to distinguish it from the propagator in the ordered phase, $g_o$, calculated in the next section.  In the following we will use standard diagrammatic techniques (see, e.g.  \cite{srednicki}).  To begin, we write down the Hamiltonian $\mathcal{H}$  defined in Eq.~\ref{eq:sphBrazo} in the main text.   Expanding  the quartic term calculated as shown in Eq.~\ref{eq:quarticintegral}, we find
 \begin{align}
\mathcal{H} & = \sum_{\bm{\ell}}  \left[ \frac{(\ell-\ell_0)^2+R^2\tau}{2} \right] | \Psi^m_{\ell}|^2   \nonumber \\ &   \quad {} +  \frac{ R^2}{3!}\sum\limits_{\{\bm{\ell}_i\}_{i=1}^3}\gamma^{(3)}\prod_{i=1}^3\Psi_{\ell_i}^{m_i} +  \frac{  R^2}{4!}\sum\limits_{\{\bm{\ell}_i\}_{i=1}^4}\gamma^{(4)}\prod_{i=1}^4\Psi_{\ell_i}^{m_i}, \label{eq:Hamiltonian} \end{align}
where we recall the definition of the bare vertex functions $\gamma^{(3)} \equiv \lambda_3( \{ \bm{\ell}_i \}_{i=1}^{3} )$ and $\gamma^{(4)} \equiv  \gamma^{(4)}( \{ \bm{\ell}_i \}_{i=1}^{4} )$ from the main text, repeated here for convenience:

\begin{align}
\gamma^{(3)} &=\lambda_3 \Upsilon_{m_1,m_2,m_3}^{\ell_1,\ell_2,\ell_3} \label{eq:bare3point} \\
\gamma^{(4)} &=\lambda_4 \sum_{\bm{\ell}} (-1)^m\Upsilon_{m_1,m_2,m}^{\ell_1,\ell_2,\ell} \Upsilon_{m_3,m_4,-m}^{\ell_3,\ell_4,\ell}  . \label{eq:bare4point}
\end{align}

We now define the Feynman rules to construct our diagrams.  The first major component  comes from the quadratic piece of the Hamiltonian, from which we derive the free propagator, denoted by a line:
\begin{align}
\parbox{12mm}{\begin{fmfgraph}(25,5) \fmfleft{v1} \fmfright{v2} \fmf{vanilla}{v1,v2}
\end{fmfgraph}}& \equiv \left\langle \Psi_{\ell_1}^{m_1} \Psi^{m_2}_{\ell_2} \right\rangle_0  =\frac{(-1)^{m_1}\delta_{\ell_1-\ell_2}\delta_{m_1+m_2}}{(\ell_1-\ell_0)^2+R^2\tau}. \label{eq:bareprop}
\end{align}
To simplify formulas that appear throughout the rest of this text, we make the definition $M(\ell,\tau)=(\ell-\ell_0)^2+\tau R^2$. The quartic term yields a fourfold vertex,
\begin{equation}
 \parbox{5mm}{\begin{fmfgraph}(12,12) \fmfleft{v1,v2} \fmfright{v3,v4} \fmf{vanilla}{v1,i1,v2} \fmf{vanilla}{v3,i1,v4} \fmfdot{i1}
\end{fmfgraph}}= -\lambda_4 R^2 \sum\limits_{\bm{\ell}}(-1)^m \Upsilon_{m_1,m_2,m}^{\ell_1,\ell_2,\ell}\Upsilon_{m_3,m_4,-m}^{\ell_3,\ell_4,\ell}=-R^2\gamma^{(4)},
\end{equation}
whereas the cubic term is denoted by
\begin{equation}
\parbox{5mm}{\begin{fmfgraph}(12,12) \fmfleft{v1,v2} \fmfright{v3} \fmf{vanilla}{v1,i1,v2} \fmf{vanilla}{v3,i1} \fmfdot{i1}
\end{fmfgraph}} =- \lambda_3 R^2 \Upsilon_{m_1,m_2,m_3}^{\ell_1,\ell_2,\ell_3}=-R^2\gamma^{(3)}.
\end{equation}
 Finally, we will sum over the angular momentum indices $\ell_i$ and $m_i$ of any internal lines (i.e., lines which connect two vertices or the same vertex to itself).  We can use these simple diagram elements to construct a perturbation expansion in the couplings $\lambda_{3,4}$, which we take to be small.

Let's begin with corrections to the inverse propagator.  Using the geometric series for the propagator \cite{srednicki}, it is possible to write the fully renormalized inverse propagator diagramatically as follows: 
\begin{equation}
g_d^{-1}(\bm{\ell}_1,\bm{\ell}_2)\equiv(\parbox{9mm}{\begin{fmfgraph}(25,5) \fmfleft{v1} \fmfright{v2} \fmf{double}{v1,v2}
\end{fmfgraph}})^{-1}= (\parbox{9mm}{\begin{fmfgraph}(25,5) \fmfleft{v1} \fmfright{v2} \fmf{vanilla}{v1,v2}
\end{fmfgraph}})^{-1}- \,\parbox{12mm}{\begin{fmfgraph*}(30,23) \fmfleft{l1} \fmfright{r1}  \fmf{plain,tension=2}{l1,v1} \fmf{phantom}{v1,i,v2} \fmfv{decor.shape=circle,decor.size=20,decor.filled=empty,label=1PI,label.dist=-8}{i} \fmf{plain,tension=2}{v2,r1}
\end{fmfgraph*}},
\end{equation}
where the fully renormalized propagator is denoted by a double line, and the second term on the right-hand side is the sum of all the two-point amputated one-particle irreducible (1PI) graphs.  These are the graphs that cannot be cut into two sub-graphs by removing a single propagator link.  There are many of these graphs that one would have to  calculate.  However, we simplify the calculation by looking at just the one-loop correction. If we include the cubic term, there are two different kinds of loop corrections: 
\begin{align}
\parbox{11mm}{\begin{fmfgraph}(30,30) \fmfleft{l1}
\fmfright{r1} \fmf{plain,tension=1}{l1,v1,l1} \fmf{plain,tension=0.6,right}{v1,v2,v1}\fmfdot{v1,v2} \fmf{plain,tension=1}{r1,v2,r1} 
\end{fmfgraph}} +
\parbox{6mm}{\begin{fmfgraph}(15,27) \fmfleft{i}
\fmfright{o} \fmf{plain}{i,v,o} \fmf{plain,tension=0.4}{v,v}\fmfdot{v}
\end{fmfgraph}}   \label{eq:disorderedloops}
\end{align}
In Brazovskii's analysis \cite{Brazovskii}, he argues that the first loop correction may be neglected relative to the second in Eq.~\ref{eq:disorderedloops} because the loop integration in the first diagram  only contributes over a narrow set of directions.  This is more difficult to see in our spherical harmonic expansion, but we may neglect this diagram in our analysis, as well.  We shall return to this point later (see Eq.~\ref{cubicloopprop}).  

 We can actually include an even larger set of diagrams if we replace the propagator in the loop with the renormalized propagator $g$ to yield a self-consistent equation:
\begin{align}
(\parbox{9mm}{\begin{fmfgraph}(25,5) \fmfleft{v1} \fmfright{v2} \fmf{double}{v1,v2}
\end{fmfgraph}})^{-1} & \approx (\parbox{9mm}{\begin{fmfgraph}(25,5) \fmfleft{v1} \fmfright{v2} \fmf{vanilla}{v1,v2}
\end{fmfgraph}})^{-1}- \parbox{6mm}{\begin{fmfgraph}(15,27) \fmfleft{i}
\fmfright{o} \fmf{plain}{i,v,o} \fmf{double,tension=0.4}{v,v}\fmfdot{v}
\end{fmfgraph}} ,\label{eq:HFProp} 
\end{align}
 where we have neglected the first loop diagram in Eq.~\ref{eq:disorderedloops} which we expect to be small.  The renormalized propagator $g$ in this approximation has a new temperature-like parameter $t_d$ (instead of $\tau$), where the subscript reminds us that we are in the disordered state.  Hence, when calculating the loop in Eq.~\ref{eq:HFProp}, we have to replace the $\tau$ in the original propagator with $t_d$.  Using our Feynman rules, this yields the following term:
 \begin{align}
\parbox{6mm}{\begin{fmfgraph}(15,27) \fmfleft{i}
\fmfright{o} \fmf{plain}{i,v,o} \fmf{double,tension=0.4}{v,v}\fmfdot{v}
\end{fmfgraph}}&=-\frac{\lambda_4 R^2}{2} \sum\limits_{\bm{\ell},\bar{\ell}}\frac{(-1)^{m_1}\Upsilon_{m,-m,0}^{\ell,\ell,\bar{\ell}}\Upsilon_{m_1,m_2,0}^{\ell_1,\ell_2,\bar{\ell}}}{M(\ell,t_d)} \label{eq:proploop0},
\end{align}
where the factor of two that appears comes from the symmetry factor of the diagram.  Using  Eq.~\ref{eq:singlemsum} to sum on $m$, we find
\begin{align}
\parbox{6mm}{\begin{fmfgraph}(15,27) \fmfleft{i}
\fmfright{o} \fmf{plain}{i,v,o} \fmf{double,tension=0.4}{v,v}\fmfdot{v}
\end{fmfgraph}}&=-\frac{\lambda_4 R^2}{2} \sum\limits_{\ell} \frac{(-1)^{m_1}(2\ell+1) \Upsilon_{m_1,m_2,0}^{\ell_1,\ell_2,0}}{\sqrt{4 \pi}\,M(\ell,t_d)} \nonumber \\
&=-\frac{\lambda_4 R^2}{8\pi} \sum\limits_{\ell}\frac{(-1)^{m_1}\delta_{\ell_1-\ell_2}\delta_{m_1+m_2}(2\ell+1)}{M(\ell,t_d)} \nonumber \\ & \equiv-\frac{ (-1)^{m_1}\delta_{\ell_1-\ell_2}\delta_{m_1+m_2}}{8\pi}\,L_1  \label{eq:proploop1}
\end{align}
where we used the special value of the Gaunt coefficient in Eq.~\ref{eq:specialGval} and identified $L_1$ as the divergent summation to be performed.

Now we must grapple with the  sum $L_1$ in Eq.~\ref{eq:proploop1}.  There is a logarithmic divergence that occurs for large $\ell$.  To remedy this divergence, we introduce a large momentum cutoff $\Lambda$.  The summation over $\ell$ may then be regularized using the Pauli-Villars technique \cite{srednicki} by introducing a modified propagator: 
\begin{equation}
g_d(\bm{\ell}_1,\bm{\ell}_2) \rightarrow \frac{(-1)^{m_1}\Lambda^2\delta_{\ell_1-\ell_2}\delta_{m_1+m_2}}{M(\ell_1,t_d)M(\ell_1,\Lambda^2/R^2)}.
\end{equation} 
Note that we will take $\Lambda \gg \ell_0$ to be very large, so that the relevant physics around $\ell \approx \ell_0$ is not modified.  With this propagator, the summation in Eq.~\ref{eq:proploop1} is convergent and, with some assistance from a computer algebra system (Mathematica v10.1, Wolfram Research, Inc., Champaign, IL), we compute
\begin{align}
\frac{L_1}{\lambda_4 R^2}&=  \ln \left(\frac{\Lambda^2}{R^2t_d} \right)+\ln (R^2 t_d)- 2 \Re\psi ^{(0)}\left(iR\sqrt{ t_d}-\ell_0 \right) \nonumber \\ & \qquad {} \qquad{}+\frac{(1+2 \ell_0)}{R\sqrt{ t_d}} \Im\psi ^{(0)}\left(i R\sqrt{t_d}-\ell_0 \right), \label{eq:L1v1}
\end{align}
where $\psi^{(0)}(z)$ is the digamma function, with properties and asymptotic expansions tabulated in  Ref.~\cite{stegun}.      We now regularize $L_1$ by subtracting off the logarithmic divergence, which in the field-theoretic language would correspond to introducing an appropriate counterterm \cite{srednicki}. Next, we assume that $\ell_0 \gg 1$, so that the argument of the digamma functions in Eq.~\ref{eq:L1v1} is large and we may make use of an asymptotic series for $\psi^{(0)}(z)$.  This yields the regularized sum
\begin{align}
L_1'& \approx   -  \lambda_4 R^2\ln \left[ 1+ \frac{\ell_0^2 }{R^2 t_d} \right] \nonumber  \\ & \qquad  +\frac{2\lambda_4 R \ell_0}{\sqrt{ t_d}} \left[   \pi \coth(\pi R \sqrt{t_d}) - \atan \left( \frac{R\sqrt{t_d}}{\ell_0}\right) \right] . \label{eq:L1v2}
\end{align}
Note that there are two important dimensionless parameters in Eq.~\ref{eq:L1v2}: $\pi R \sqrt{t_d}$ and $R \sqrt{t_d}/\ell_0$.  When we take the $R \rightarrow \infty$ limit, we want to be sure to recover the correct planar limit described by the original Brazovskii analysis (adapted to two dimensions) \cite{Brazovskii}.  To do this, we must take $\ell_0 \rightarrow \infty$ as $R \rightarrow \infty$ such that $\ell_0/R = k_0$ remains constant.  Recall that $k_0=2\pi/\lambda_0$ corresponds to the special wavevector associated with the unstable wavelength $\lambda_0$.  Moreover, since we are interested in small $t_d$  where we find the largest contributions from fluctuations, we may approximate $L_1'$ by 
\begin{equation}
L_1' \approx\frac{2\pi  \lambda_4 R\ell_0}{\sqrt{ t_d}} \, \coth(\pi R \sqrt{t_d}). \label{eq:L1final}
\end{equation}
Substituting Eq.~\ref{eq:L1final} into Eq.~\ref{eq:proploop1} and evaluating the latter equation at $\ell_1=\ell_2=\ell_0$  yields the self-consistent equation  for $t_d$ in the main text (Eq.~\ref{eq:propcrossover}) via Eq.~\ref{eq:HFProp}.  Alternatively, Eq.~\ref{eq:HFProp} may be written  as an equation for the fluctuation-renormalized propagator $g_d \equiv g_d(\bm{\ell}_1,\bm{\ell}_2)$.  Note that this propagator is diagonal, i.e., it vanishes unless $\ell_1=\ell_2$ and $m_1=-m_2$:
\begin{equation}
g_d^{-1} =\left[ M(\ell_1,\tau) +\frac{L_1' }{8\pi} \right](-1)^{m_1} \delta_{m_1+m_2} \delta_{\ell_1-\ell_2}, \label{eq:HFProp2}
\end{equation}
where $L_1'$ is given in Eq.~\ref{eq:L1final}.

The vertex function $\Gamma^{(4)}$ is calculated in a similar way. As discussed in the main text, we are only interested in the quartic term corrections (the $\lambda_3=0$ case).  This time, there are three relevant amputated diagrams:
\begin{align}
\Gamma^{(4)} & = \parbox{5mm}{\begin{fmfgraph}(12,12) \fmfleft{v1,v2} \fmfright{v3,v4} \fmf{vanilla}{v1,i1,v2} \fmf{vanilla}{v3,i1,v4} \fmfdot{i1}
\end{fmfgraph}} -\parbox{8mm}{\begin{fmfgraph}(20,20) \fmfleft{l1,l2}
\fmfright{r1,r2} \fmf{plain,tension=3}{l1,v1,l2} \fmf{double,tension=0.4,right}{v1,v2,v1}\fmfdot{v1,v2} \fmf{plain,tension=3}{r1,v2,r2}
\end{fmfgraph}}  - \parbox{8mm}{\begin{fmfgraph}(20,20) \fmftop{l1,l2}
\fmfbottom{r1,r2} \fmf{plain,tension=3}{l1,v1,l2} \fmf{double,tension=0.4,right}{v1,v2,v1}\fmfdot{v1,v2} \fmf{plain,tension=3}{r1,v2,r2} \end{fmfgraph}} - \parbox{8mm}{\begin{fmfgraph}(20,20) \fmftop{l1,l2}
\fmfbottom{r1,r2} \fmf{plain,tension=0}{l1,v1,r2} \fmf{double,tension=0.4,right}{v1,v2,v1}\fmfdot{v1,v2} \fmf{plain,tension=0}{r1,v2,l2}  \fmf{phantom,tension=3}{l1,v1,l2}  \fmf{phantom,tension=3}{r1,v2,r2}  \end{fmfgraph}}.
\end{align}
Let's compute the first one as the rest are similar. We have
\begin{align}
\parbox{8mm}{\begin{fmfgraph}(20,20) \fmfleft{l1,l2}
\fmfright{r1,r2} \fmf{plain,tension=3}{l1,v1,l2} \fmf{double,tension=0.4,right}{v1,v2,v1} \fmf{plain,tension=3}{r1,v2,r2} \fmfdot{v1,v2}
\end{fmfgraph}} & = \frac{R^4\lambda_4^2}{2} \sum_{\bm{\ell}_5,\bm{\ell}_6,\bar{\bm{\ell}},\bar{\bm{\ell}}'} (-1)^{\bar{m}+\bar{m}'}\Upsilon_{m_1,m_2,\bar{m}}^{\ell_1,\ell_2,\bar{\ell}}\Upsilon_{m_5,m_6,-\bar{m}}^{\ell_5,\ell_6,\bar{\ell}} \nonumber\\ & \qquad {} \times \frac{(-1)^{m_6+m_5} \Upsilon_{-m_5,-m_6,\bar{m}'}^{\ell_5,\ell_6,\bar{\ell}'}\Upsilon_{m_3,m_4,-\bar{m}'}^{\ell_3,\ell_4,\bar{\ell}'}}{M(\ell_5,t_d)M(\ell_6,t_d)} \nonumber \\ & =  \frac{\lambda_4 R^2}{2} \sum_{\ell_5,\ell_6,\bar{\bm{\ell}}} (-1)^{\bar{m}}\Upsilon_{m_1,m_2,\bar{m}}^{\ell_1,\ell_2,\bar{\ell}}\Upsilon_{m_3,m_4,-\bar{m}}^{\ell_3,\ell_4,\bar{\ell}} \Pi(\bar{\ell}), \label{eq:vertexloop0}
\end{align}
where $\bm{\ell}_{1,2,3,4}$ are the indices of the four external (amputated) legs. We have performed the summations over $m_{5,6}$ using Eq.~\ref{eq:sumtwoms} and identified our loop summation
\begin{align}
\Pi(\ell) & =  \sum_{\ell_1,\ell_2} \frac{\lambda_4 R^2\Upsilon_{0,0,0}^{\ell_1,\ell_2,\ell}}{\sqrt{4\pi(2\ell+1)}} \prod_{i=1}^2 \frac{\sqrt{2\ell_i+1}}{M(\ell_i,t_d)}. \label{eq:vertexloop1}
\end{align}

The most divergent contribution to the sums over $\ell_{1,2}$ in Eq.~\ref{eq:vertexloop1} comes from $\ell_1 \approx \ell_2 \approx \ell_0$.  The Gaunt coefficient in Eq.~\ref{eq:vertexloop1} contains no divergences, so we will set $\ell_1=\ell_2=\ell_0$ in this coefficient.  This leaves us with the single sum
\begin{equation}
\Pi(\ell) \propto   L_2 \equiv  \sum_{\bar{\ell}}  \frac{\lambda_4 R^2(2\bar{\ell}+1)}{[M(\bar{\ell},t_d)]^2}, \label{eq:vertexloop2}
\end{equation}
where the constant of proportionality is easily read off from Eq.~\ref{eq:vertexloop1}. The sum $L_2$ in Eq.~\ref{eq:vertexloop2} does not need regularization and reads
\begin{align}
L_2 & =  \frac{\lambda_4 \ell_0 }{R\sqrt{t_d}} \Bigg[\frac{ \Im  \psi ^{(0)}(\rho )}{t_d}  - R^2\Im \psi ^{(1)}( \rho) \nonumber \\ & \qquad \quad \quad {}   -\frac{R\Re \psi^{(1)}(\rho )}{\sqrt{t_d}} \Bigg]   ,
\end{align}
where $\rho \equiv iR \sqrt{t_d} -\ell_0$ and  $\psi^{(1)}(z)$ is the first derivative of the digamma function.  Although we do not have to regularize, we will want to capture the correct asymptotic behavior of the sum $L_2$. Once again, we are interested in the two limits $R \sqrt{t_d} \rightarrow 0$ and $R \sqrt{t_d} \rightarrow \infty$ in such a way that $\ell_0/R$ remains constant.  Once again making use of the asymptotic properties of the polygamma functions \cite{stegun}, we find
\begin{equation}
L_2  \approx \frac{\lambda_4\pi^2\ell_0}{t_d \sinh^2(\pi R \sqrt{t_d})}+ \frac{ \lambda_4\ell_0\pi \coth(\pi R \sqrt{t_d})}{R t_d^{3/2}} , \label{eq:L2final}
\end{equation}
which manifestly yields the vertex function result discussed in the main text.  The sum $L_2$ also clearly diverges in the small $t_d$ limit, either as $t_d^{-3/2}$ in the planar limit $(R \sqrt{t_d} \rightarrow \infty$ with $\ell_0/R$ fixed) or as $t_d^{-2}$ in the finite size scaling regime $(R \sqrt{t_d} \rightarrow 0)$.

Finally, let us return briefly to our neglected loop correction to the propagator.  Now that we have calculated $L_2$, we may use the same calculation to evaluate the following diagram, which also includes a summation over two propagators:
\begin{align}
\parbox{11mm}{\begin{fmfgraph}(30,30) \fmfleft{l1}
\fmfright{r1} \fmf{plain,tension=1}{l1,v1,l1} \fmf{double,tension=0.6,right}{v1,v2,v1}\fmfdot{v1,v2} \fmf{plain,tension=1}{r1,v2,r1} 
\end{fmfgraph}} & = \frac{R^4 \lambda_3^2}{4 \sqrt{\pi}} \,(-1)^{m_1}\delta_{m_1+m_2}\delta_{\ell_1-\ell_2} \nonumber \\ & \quad {} \times\sum_{\bar{\ell}_1,\bar{\ell}_2} \frac{\Upsilon_{0,0,0}^{\bar{\ell}_1,\bar{\ell}_2,\ell_1}}{\sqrt{2\ell_1+1}} \prod_{i=1}^2 \frac{\sqrt{2\bar{\ell}_i+1}\,}{M(\bar{\ell}_i,t_d)}. \label{cubicloopprop}
\end{align}
So, as before, we look at the most divergent contribution which occurs when $\bar{\ell}_{1} = \bar{\ell}_2 \approx \ell_0$.  We are again left a single summation which gives us the same divergences as Eq.~\ref{eq:L2final}.  Therefore, at our momenta of interest $\ell_1 =\ell_0$, we find that when $R \sqrt{t_d} \gg 1$, the diagram scales like $\Upsilon^{\ell_0,\ell_0,\ell_0}_{0,0,0}R\lambda_3^2 t_d^{-3/2}\sqrt{\ell_0}$ and like $\Upsilon^{\ell_0,\ell_0,\ell_0}_{0,0,0}R^2 \lambda_3^2 t_d^{-2}\sqrt{\ell_0}$ when $R \sqrt{t_d} \ll 1$.  In either case, when $\ell_0 \gg 1$, these contributions are much smaller than the loop correction we already calculated in Eq.~\ref{eq:proploop1} because $\Upsilon^{\ell_0,\ell_0,\ell_0}_{0,0,0} \sqrt{\ell_0} \approx \mathrm{const.}$ for large $\ell_0$,  whereas the contribution in Eq.~\ref{eq:proploop1} increases linearly with $\ell_0$.  Hence, just as in the Brazovskii analysis, we may neglect this loop correction when $\ell_0 \gg 1$.
 
\section{The Ordered State and $\Delta \Phi$ \label{appxordered}}

In this Appendix, we calculate the free energy change $\Delta \Phi$ between the disordered state and the ordered one. We'll also develop our perturbation theory around the ordered state $\bar{\Psi}$.
Recall that in the ordered state, we have to expand around a new potential minimum, so that our Hamiltonian has a different form and a different set of Feynman rules.  First, instead of the fields $\Psi_{\ell_i}^{m_i}$, our new fluctuating fields are the modes  $\psi_{\ell_i}^{m_i}$ of the fluctuations $\psi$ away from the ordered state $\bar{\Psi}$.  The Hamiltonian for these fluctuating modes includes all of the terms in the Hamiltonian in Eq.~\ref{eq:Hamiltonian}.  However, there are new cross terms coming from powers of the expanded modes $\Psi^{m_i}_{\ell_i} =\bar{\Psi}_{\ell_i}^{m_i} + \psi_{\ell_i}^{m_i}$, which we will denote by $\Delta \mathcal{H}$.  These new terms are all nonlinear in  $\psi^{m_i}_{\ell_i}$. The fields $\psi$ describe fluctuations away from the potential minimum.  So, we have  
\begin{align}
\Delta \mathcal{H} = &\frac{ R^2}{6} \sum_{\{\bm{\ell}_i \}_{i=1}^4 } \gamma^{(4)} \psi_{\ell_1}^{m_1}   \psi_{\ell_2}^{m_2}   \left[   \psi_{\ell_3}^{m_3}  +\frac{3}{2}\bar{\Psi}^{m_3}_{\ell_3}   \right]\bar{\Psi}^{m_4}_{\ell_4} . \label{eq:ordereddeltaH}\\
&+\frac{R^2}{2}\sum_{\{\bm{\ell}_i \}_{i=1}^3 }\gamma^{(3)}  \psi_{\ell_1}^{m_1}       \psi_{\ell_2}^{m_2} \bar{\Psi}^{m_3}_{\ell_3} \nonumber
\end{align}
Note that we have also ignored all the terms that do not depend on $\psi$, as these do not contribute to any correlation functions of the $\psi$ fields. These new terms introduce three new kinds of vertices, with three or two legs which we may contract.  We denote these vertices as follows:
\begin{align}
 \parbox{6mm}{\begin{fmfgraph}(15,15) \fmfleft{v1,v2} \fmfright{v3,v4} \fmf{vanilla}{v1,i1} \fmf{vanilla}{i1,v2} \fmf{vanilla}{v3,i1,v4} \fmfdot{i1} \fmfv{decor.shape=circle,decor.filled=empty,decor.size=4}{v2}
\end{fmfgraph}} &= -\frac{ R^2}{6} \sum_{\bm{\ell}_4}\gamma^{(4)}(\bm{\ell}_1,\bm{\ell}_2,\bm{\ell}_3,\bm{\ell}_4)\bar{\Psi}^{m_4}_{\ell_4} \label{eq:threepointordered}\\
 \parbox{6mm}{\begin{fmfgraph}(15,15) \fmfleft{v1,v2} \fmfright{v3,v4} \fmf{vanilla}{v1,i1} \fmf{vanilla}{i1,v2} \fmf{vanilla}{v3,i1} \fmf{vanilla}{i1,v4} \fmfdot{i1} \fmfv{decor.shape=circle,decor.filled=empty,decor.size=4}{v2,v4}
\end{fmfgraph}} &= -\frac{ R^2}{2} \sum\limits_{\bm{\ell}_3,\bm{\ell}_4}\gamma^{(4)}(\bm{\ell}_1,\bm{\ell}_2,\bm{\ell}_3,\bm{\ell}_4)\bar{\Psi}^{m_3}_{\ell_3}\bar{\Psi}^{m_4}_{\ell_4}, \label{eq:twopointordered}\\
\parbox{6mm}{\begin{fmfgraph}(15,20) \fmfleft{v1} \fmfright{v3} \fmftop{v2} \fmf{vanilla}{v1,i1} \fmf{vanilla}{i1,v2} \fmf{vanilla}{v3,i1} \fmf{vanilla,tension=0}{i1,v2} \fmfdot{i1} \fmfv{decor.shape=circle,decor.filled=empty,decor.size=4}{v2}
\end{fmfgraph}} &=-\frac{R^2}{2}\sum_{\bm{\ell}_3 }\gamma^{(3)} (\bm{\ell}_1,\bm{\ell}_2,\bm{\ell}_3)   \bar{\Psi}^{m_3}_{\ell_3}  \label{eq:twopointordered2}
\end{align}
where the circles on the legs indicate the insertion of an ordered field mode $\bar{\Psi}^{m_i}_{\ell_i}$.  Note that all of our ordered fields will have $\ell_i=\ell_0$, so we may omit the index $\ell$ of these modes in the following.  When calculating averages of the fields $\psi$, these two new vertices must be included in the Feynman rules already defined in the previous section.

The vertex in Eq.~\ref{eq:threepointordered}  is the next-lowest order contribution to the three-point function $\langle \psi \psi \psi \rangle$ (after the bare contribution from the cubic term which vanishes for any $\ell_0 >0$, anyway), whereas the vertex in Eq.~\ref{eq:twopointordered} contributes a new term to the propagator equation.    Before calculating any loop corrections, let's study the scaling properties of these two new vertices for small $\lambda_3$.  Recall from the main text that the ordered state amplitude $\bar{a}$ satisfies $\bar{a}^2 \approx 4 \pi t_o/\lambda_4 $ when $\lambda_3=0$ (see also Eq.~\ref{eq:orderedstatesol} below).  Hence, the circles in the new vertices will bring in scaling factors of   $\bar{\Psi}\sim \sqrt{t_o/\lambda_4 }$ (although this scaling may be complicated by the presence of the cubic term).  Then, we may verify that the contribution from the three-point function  $\lambda_4 \langle \psi \psi \psi \rangle$ is small relative to the two point function contribution $\lambda_4 \langle \psi \psi \rangle \bar{\Psi}$:  $\langle \psi \psi \psi \rangle/(\langle \psi \psi \rangle \bar{\Psi})\sim \sqrt{\lambda_4 t_o}/( t_o \sqrt{t_o/\lambda_4}) \sim \lambda_4/t_o \ll 1$.  It is possible that this particular scaling fails if the cubic coupling $\lambda_3$ is sufficiently large.  We still expect to be able to neglect the three-point function, because both the leading order contribution to $\langle \psi \psi \psi \rangle$ and $\langle \psi \psi \rangle \bar{\Psi}$ are proportional to the  ordered state amplitude  within our approximation, so the three-point function should still be relatively small.  However,   a detailed check  is beyond the scope of this analysis.  So, following Brazovskii, we now neglect the three-point function contribution and calculate the equation for the magnetic field $h$: 
  \begin{align}
h_{m} &\ =\frac{1}{4\pi R^2} \Big\langle \frac{\delta \mathcal{H}[\psi+\bar{\Psi}]}{\delta \bar{\Psi}_{m}} \Big\rangle \nonumber \\ & = \frac{1}{4\pi R^2} \left[\Big\langle \frac{\delta ( \mathcal{H}[\bar{\Psi}])}{\delta \bar{\Psi}_{m}} \Big\rangle + \Big\langle \frac{\delta (\Delta \mathcal{H})}{\delta \bar{\Psi}_{m}} \Big\rangle  \right] \nonumber   \\
&    =    \frac{\tau \bar{\Psi}^*_{m}}{4\pi} + \frac{\lambda_4 }{8\pi} \sum_{\bm{\ell}_{1,2},m_3,\bar{\bm{\ell}}} (-1)^{\bar{m}}\Upsilon^{\ell_1,\ell_2,\bar{\ell}}_{m_1,m_2 ,\bar{m}}   \Upsilon^{\ell_0,\ell_0,\bar{\ell}}_{m_3,m,-\bar{m}}  \nonumber \\
&  \qquad \quad  \times \bar{\Psi}_{m_3}  \left[ \langle \psi_{\ell_1}^{m_1}  \psi_{\ell_2}^{m_2} \rangle+ \frac{\delta_{\ell_2-\ell_0}\delta_{\ell_1-\ell_0}}{3}   \bar{\Psi}_{m_1} \bar{\Psi}_{m_2} \right]  \nonumber \\ & \qquad {}+ \frac{\lambda_3}{8 \pi} \sum_{m_{1,2}} \Upsilon_{m,m_1,m_2}^{\ell_0,\ell_0,\ell_0} \bar{\Psi}_{m_1} \bar{\Psi}_{m_2} , \label{eq:generalh0} 
\end{align}
where in the second line we retain just the terms in the Hamiltonian expanded around the ordered state, $\mathcal{H}[\psi+\bar{\Psi}]$, which retain at least a single power of $\bar{\Psi}$, since we take a functional derivative with respect to the ordered state modes $\bar{\Psi}_m$.  We also drop all terms that are linear in the fluctuations $\psi$, since $\langle \psi \rangle =0$ as discussed in the main text.   Our task now is to write $h_m$ \textit{just} in terms of the ordered state modes $\bar{\Psi}_m$ and the renormalized value of $\tau$ in the ordered state, $t_o$.  Before proceeding, we will make an approximation (partially justified below) that  only the diagonal components $m_1=-m_2=m$ and $\ell_1=\ell_2=\ell$ contribute to the two-point function $g_o(\bm{\ell}_1,\bm{\ell}_2)= \langle \psi_{\ell_1}^{m_1} \psi_{\ell_2}^{m_2}\rangle$.  This is manifestly true for the disordered state, as can be seen from Eq.~\ref{eq:HFProp2}.  In this diagonal approximation, Eq.~\ref{eq:generalh0} reduces to
 \begin{align}
 h_{m} &   =    \frac{\tau \bar{\Psi}^*_{m}}{4\pi} +\frac{\lambda_4 }{8\pi}  \sum_{\bm{\ell}_1,\bar{\bm{\ell}}} \Upsilon^{\ell_1,\ell_1,\bar{\ell}}_{m_1,-m_1 ,0}   \Upsilon^{\ell_0,\ell_0,\bar{\ell}}_{m,-m,0}\langle \psi_{\ell_1}^{m_1}  \psi_{\ell_1}^{-m_1} \rangle\bar{\Psi}_{-m}  \nonumber \\
&  {} +\frac{\lambda_4 }{24\pi} \sum_{m_{1,2,3},\bar{\bm{\ell}}} (-1)^{\bar{m}}\Upsilon^{\ell_0,\ell_0,\bar{\ell}}_{m_1,m_2 ,\bar{m}}   \Upsilon^{\ell_0,\ell_0,\bar{\ell}}_{m_3,m,-\bar{m}}   \prod_{i=1}^3  \bar{\Psi}_{m_i} \nonumber \\
&  {}+ \frac{\lambda_3}{8 \pi} \sum_{m_{1,2}} \Upsilon_{m,m_1,m_2}^{\ell_0,\ell_0,\ell_0} \bar{\Psi}_{m_1} \bar{\Psi}_{m_2}. \label{eq:generalh}
 \end{align}   

   Our equation of state, Eq.~\ref{eq:generalh},  depends only on the two-point function $g_o(\bm{\ell}_1,\bm{\ell}_2)\equiv \left\langle \psi^{m_1}_{\ell_1} \psi^{m_2}_{\ell_2} \right\rangle$ of fluctuations in the ordered state.  To calculate this function in the Hartree-Fock approximation, we proceed as in the disordered state calculation and construct a diagrammatical equation:
\begin{align}
(\parbox{9mm}{\begin{fmfgraph}(25,5) \fmfleft{v1} \fmfright{v2} \fmf{double}{v1,v2}
\end{fmfgraph}})^{-1}  &\approx (\parbox{9mm}{\begin{fmfgraph}(25,5) \fmfleft{v1} \fmfright{v2} \fmf{vanilla}{v1,v2}
\end{fmfgraph}})^{-1}- \left[\parbox{6mm}{\begin{fmfgraph}(15,15) \fmfleft{v1} \fmfright{v3} \fmftop{v2} \fmf{vanilla}{v1,i1} \fmf{vanilla,tension=0}{i1,v2} \fmf{vanilla}{v3,i1}  \fmfdot{i1} \fmfv{decor.shape=circle,decor.filled=empty,decor.size=4}{v2}
\end{fmfgraph}} +\begin{fmfgraph}(15,10) \fmfleft{v1,v2} \fmfright{v3,v4} \fmf{vanilla}{v1,i1} \fmf{vanilla,tension=0}{i1,v2} \fmf{vanilla}{v3,i1} \fmf{vanilla,tension=0}{i1,v4} \fmfdot{i1} \fmfv{decor.shape=circle,decor.filled=empty,decor.size=4}{v2,v4}
\end{fmfgraph} \right] \nonumber \\ & \qquad \qquad {}   - \left[  \parbox{6mm}{\begin{fmfgraph}(15,27) \fmfleft{i}
\fmfright{o} \fmf{plain}{i,v,o} \fmf{double,tension=0.4}{v,v}\fmfdot{v}
\end{fmfgraph}} + \parbox{8mm}{\begin{fmfgraph}(20,20) \fmfleft{l1,l2}
\fmfright{r1,r2} \fmf{plain,tension=3}{l1,v1,l2} \fmf{double,tension=0.4,right}{v1,v2,v1}\fmfdot{v1,v2} \fmf{plain,tension=3}{r1,v2,r2} \fmfv{decor.shape=circle,decor.filled=empty,decor.size=4}{l2,r2}
\end{fmfgraph}}    + \parbox{11mm}{\begin{fmfgraph}(30,20) \fmfleft{l1,l2}
\fmfright{r1} \fmf{plain,tension=3}{l1,v1,l2} \fmf{double,tension=0.7,right}{v1,v2,v1}\fmfdot{v1,v2} \fmf{plain,tension=2}{r1,v2} \fmfv{decor.shape=circle,decor.filled=empty,decor.size=4}{l2}
\end{fmfgraph}}    \right],\label{eq:HFPropordered} 
\end{align} 
where the double line now indicates a propagator with the ordered state temperature parameter $t_o$.  Like the disordered state version, the parameter $t_o$ will be independent of the mode indices $\ell_{1,2}$ and $m_{1,2}$.  This ``isotropic'' approximation, however, must be justified as the ordered state corrections include new terms (not present in the disordered state calculation in Eq.~\ref{eq:HFProp})  with non-trivial $m$ dependence. First, there are two new diagrams without any loops:
\begin{align} 
\parbox{6mm}{\begin{fmfgraph}(15,20) \fmfleft{v1} \fmfright{v3} \fmftop{v2} \fmf{vanilla}{v1,i1} \fmf{vanilla,tension=0}{i1,v2} \fmf{vanilla}{v3,i1}  \fmfdot{i1} \fmfv{decor.shape=circle,decor.filled=empty,decor.size=4}{v2}
\end{fmfgraph}}& = - \frac{\lambda_3 R^2}{2} \sum_{m_3} \Upsilon^{\ell_1,\ell_2,\ell_0}_{m_1,m_2,m_3} \bar{\Psi}_{m_3} \label{eq:treediagramcubic}\\
\begin{fmfgraph}(15,10) \fmfleft{v1,v2} \fmfright{v3,v4} \fmf{vanilla}{v1,i1} \fmf{vanilla,tension=0}{i1,v2} \fmf{vanilla}{v3,i1} \fmf{vanilla,tension=0}{i1,v4} \fmfdot{i1} \fmfv{decor.shape=circle,decor.filled=empty,decor.size=4}{v2,v4}
\end{fmfgraph} & =  - \frac{\lambda_4 R^2}{2} \sum_{\bar{\bm{\ell}},m_{3,4}} (-1)^{\bar{m}}\Upsilon^{\ell_1,\ell_2,\bar{\ell}}_{m_1,m_2,\bar{m}} \nonumber \\ &   \qquad \qquad \qquad \times \Upsilon^{\ell_0,\ell_0,\bar{\ell}}_{m_3,m_4,-\bar{m}}\bar{\Psi}_{m_3} \bar{\Psi}_{m_4}, \label{eq:treediagram}
\end{align}
where the external legs have indices $\bm{\ell}_{1,2}$.  This contribution is called the ordered state term in the main text (see Eq.~\ref{eq:orderedprop}).  As usual, this contribution will be important for the special modes with $\ell_1=\ell_2=\ell_0$. A scaling analysis at $\lambda_3=0$ reveals that Eq.~\ref{eq:treediagram} is the most important difference between the propagators in the ordered and disordered states.  The contribution from Eq.~\ref{eq:treediagram} scales like $\lambda_4 R^2 \bar{a}^2 \sim R^2t_o$ due to the presence of the ordered state legs.  The loop corrections scale  like $\lambda_4\ell_0 R t_o^{-1/2}$ for the planar limit $R \sqrt{t_o} \gg 1 $ and   $  \lambda_4\ell_0 t_o^{-1} $  for the finite size scaling regime $R \sqrt{t_o} \ll 1 $ .  So, loop corrections are suppressed by the coupling constant $\lambda_4$ relative to the correction without any loops, and the latter is the largest correction in this perturbative analysis.  As discussed in more detail below, we expect a similar suppression when $\lambda_3 \neq 0$, but will make no detailed checks.

The cubic term, Eq.~\ref{eq:treediagramcubic}, also contributes.  However, note that by the property of the Gaunt coefficients, it only contributes for a single, special non-zero ordered state mode $\bar{\Psi}_m$ with $m=m_1+m_2$.  Conversely, the term in Eq.~\ref{eq:treediagram} will have contributions from all ordered state modes.   So,  we will neglect this cubic term contribution for now, and then check that this is reasonable approximation within our isotropic approximation (see Eq.~\ref{eq:cubicvanishes}).  The same argument applies for the last loop correction in Eq.~\ref{eq:HFPropordered}, which is also generated by the cubic term.  We expect it to be negligible relative to the other loop contributions.  For now, we focus on the contribution in Eq.~\ref{eq:treediagram}.

For ordered states with a single mode, $\bar{\Psi}_m$, the contribution in Eq.~\ref{eq:treediagram} vanishes except when $m_1=-m_2$.  We also expect  terms with $m_1 \neq -m_2$ to be suppressed because, in the absence of a cubic term, they will only contribute when they satisfy the sum rule $m_1+m_2+m_3+m_4=0$  where $m_{3,4}$ are indices which contribute to the ordered state $\bar{\Psi}$.     So, we  assume that our ordered state propagator is diagonal, i.e., vanishes whenever $m_1 \neq -m_2$. This approximation has an analogy in the Brazovskii analysis: The propagator corrections with external momenta not adding up to zero (pointing in opposite directions) are thrown out. So, our contribution of interest is
\begin{align}
\, \, \,\,\begin{fmfgraph*}(15,10) \fmfleft{v1,v2} \fmfright{v3,v4} \fmf{vanilla}{v1,i1} \fmf{vanilla,tension=0}{i1,v2} \fmf{vanilla}{v3,i1} \fmf{vanilla,tension=0}{i1,v4} \fmfdot{i1} \fmfv{decor.shape=circle,decor.filled=empty,decor.size=4}{v2,v4}
\fmfv{label.dist=-1,label=$\scriptstyle\ell_0,,m_1$}{v1}  \fmfv{label.dist=-0.8,label=$\scriptstyle\ell_0,,-m_1$}{v3}  \end{fmfgraph*} \quad \quad \, & \approx - \frac{\lambda_4 R^2 (-1)^{m_1}}{8 \pi} \sum_{m}\alpha_{m_1,m}^{\ell_0}|\bar{\Psi}_{m}|^2, \label{eq:treediagram2}
\end{align}  
where we have indicated the appropriate mode indices on the external legs  and introduced an important combination of Gaunt coefficients:
 \begin{equation}
 \alpha_{m_1,m_2}^{\ell_0}=4\pi\sum_{\bar{\ell}}(-1)^{m_1+m_2}\Upsilon^{\ell_0,\ell_0,\bar{\ell}}_{m_1,-m_1,0}\Upsilon^{\ell_0,\ell_0,\bar{\ell}}_{m_2,-m_2,0}. \label{eq:alphadef}
 \end{equation}   Let us now move on to the loop corrections.

The first loop correction in Eq.~\ref{eq:HFPropordered} is the same Hartree-Fock contribution we found for the disordered state in Eq.~\ref{eq:proploop1}.  So, there is nothing new here except for a replacement of $t_d$ by $t_o$.  However, we may rewrite the contribution in a convenient way as follows:
\begin{equation}
 \parbox{6mm}{\begin{fmfgraph}(15,27) \fmfleft{i}
\fmfright{o} \fmf{plain}{i,v,o} \fmf{double,tension=0.4}{v,v}\fmfdot{v}
\end{fmfgraph}} = -\frac{ R^2}{2} \sum\limits_{\bm{\ell}_{3,4}} \gamma^{(4)}(\bm{\ell}_1,\bm{\ell}_2,\bm{\ell}_3,\bm{\ell}_4)g_o(\bm{\ell}_3,\bm{\ell}_4). \label{eq:generalloop}
\end{equation}
The first new loop contribution in the ordered state is reminiscent of the $\Gamma^{(4)}$ loop correction in the disordered state (Eq.~\ref{eq:vertexloop0}):\begin{align}
\parbox{8mm}{\begin{fmfgraph}(20,20) \fmfleft{l1,l2}
\fmfright{r1,r2} \fmf{plain,tension=3}{l1,v1,l2} \fmf{double,tension=0.4,right}{v1,v2,v1}\fmfdot{v1,v2} \fmf{plain,tension=3}{r1,v2,r2} \fmfv{decor.shape=circle,decor.filled=empty,decor.size=4}{l2,r2}
\end{fmfgraph}}   & = \frac{R^4\lambda_4^2}{2} \sum_{\bar{\ell}_{1,2},m_{3,4}} \frac{ (-1)^{\bar{m}}\Upsilon_{m_1,m_3,\bar{m}}^{\ell_1,\ell_0,\bar{\ell}}\Upsilon_{m_4,m_2,-\bar{m}}^{\ell_0,\ell_2,\bar{\ell}} }{M(\bar{\ell}_1,t_o)M(\bar{\ell}_2,t_o)} \nonumber \\ & \qquad  \times  \bar{\Psi}_{m_3} \bar{\Psi}_{m_4} \Upsilon_{0,0,0}^{\bar{\ell}_1,\bar{\ell}_2, \bar{\ell}} \sqrt{\frac{(2\bar{\ell}_1+1)(2 \bar{\ell}_2+1)}{4\pi(2\bar{\ell}+1)}}. \label{eq:newvertexloop}
\end{align}

We will now explicitly show that this loop correction is negligible compared to Eq.~\ref{eq:generalloop} when $\lambda_3=0$.  First, we look at the largest contribution from this term, which happens near the region $\bar{\ell}_{1,2} \approx \ell_0$ in Eq.~\ref{eq:newvertexloop}.  As discussed in the main text and above, we neglect the off-diagonal contributions to the two-point function, so we may set the external leg indices  $\bm{\ell}_{1,2}$ to $\ell_1=\ell_2 = \ell_0$ and $m_1=-m_2$.  We then find an expression similar to the one for the vertex correction in Eqs.~\ref{eq:vertexloop0}, \ref{eq:vertexloop2}:
\begin{align}
\parbox{8mm}{\begin{fmfgraph*}(20,20) \fmfleft{l1,l2}
\fmfright{r1,r2} \fmf{plain,tension=3}{l1,v1,l2} \fmf{double,tension=0.4,right}{v1,v2,v1} \fmfdot{v1,v2} \fmf{plain,tension=3}{r1,v2,r2}  \fmfv{decor.shape=circle,decor.filled=empty,decor.size=4}{l2,r2} \fmfv{label.dist=-1,label=$\scriptstyle  m_1$}{l1}  \fmfv{label.dist=-0.8,label=$\scriptstyle -m_1$}{r1}
\end{fmfgraph*}}   & = \frac{(-1)^{m_1}R^4\lambda_4^2}{2} \sum_{m_2,\bar{\ell}}  \frac{A_{\ell_0}(m_1,m_2) | \bar{\Psi}_{m_2}|^2(2\bar{\ell}+1)}{[M(\bar{\ell},t_o)]^2} \nonumber  \\
& =\frac{(-1)^{m_1} \pi R^2 \ell_0\lambda_4^2}{2} \sum_{m_2} A_{\ell_0}(m_1,m_2) | \bar{\Psi}_{m_2}|^2  \nonumber \\ & \qquad \times\left[ \frac{\pi }{t_o \sinh^2(\pi R \sqrt{t_o})}+ \frac{  \coth(\pi R \sqrt{t_o})}{ Rt_o^{3/2}} \right], 
\label{eq:newvertexloop2}
\end{align}  
where we have indicated the appropriate $m$ indices on the external legs of the diagram.  This contribution includes a special function $A_{\ell_0}(m_1,m_2)$ that introduces an $m$-dependence to the diagram:
\begin{align}
A_{\ell_0}(m_1,m_2)\equiv \sum_{m_2,\bm{\ell}} \frac{ [\Upsilon_{m_1,m_2,m}^{\ell_0,\ell_0,\ell}]^2 \Upsilon_{0,0,0}^{\ell_0,\ell_0 ,\ell}}{\sqrt{4\pi(2\ell+1)}}.
\end{align}
Recognizing that $|\bar{\Psi}_{m_2}|^2 \sim \bar{a}^2 \sim t_o/\lambda_4$ in Eq.~\ref{eq:newvertexloop2}, it is easy to see that this new loop correction scales in the same way as the loop correction in Eq.~\ref{eq:generalloop}.  So, we will have to analyze the function $A_{\ell_0}$ in some detail to prove that, much like in the Brazovskii analysis, Eq.~\ref{eq:newvertexloop2} contributes significantly only for special values of $m_1$: when $m_1$ is equal to one of the $m$'s that contributes to the ordered state $\bar{\Psi}$.

 \begin{figure}[t]
\centerline{\includegraphics[height=3.5in]{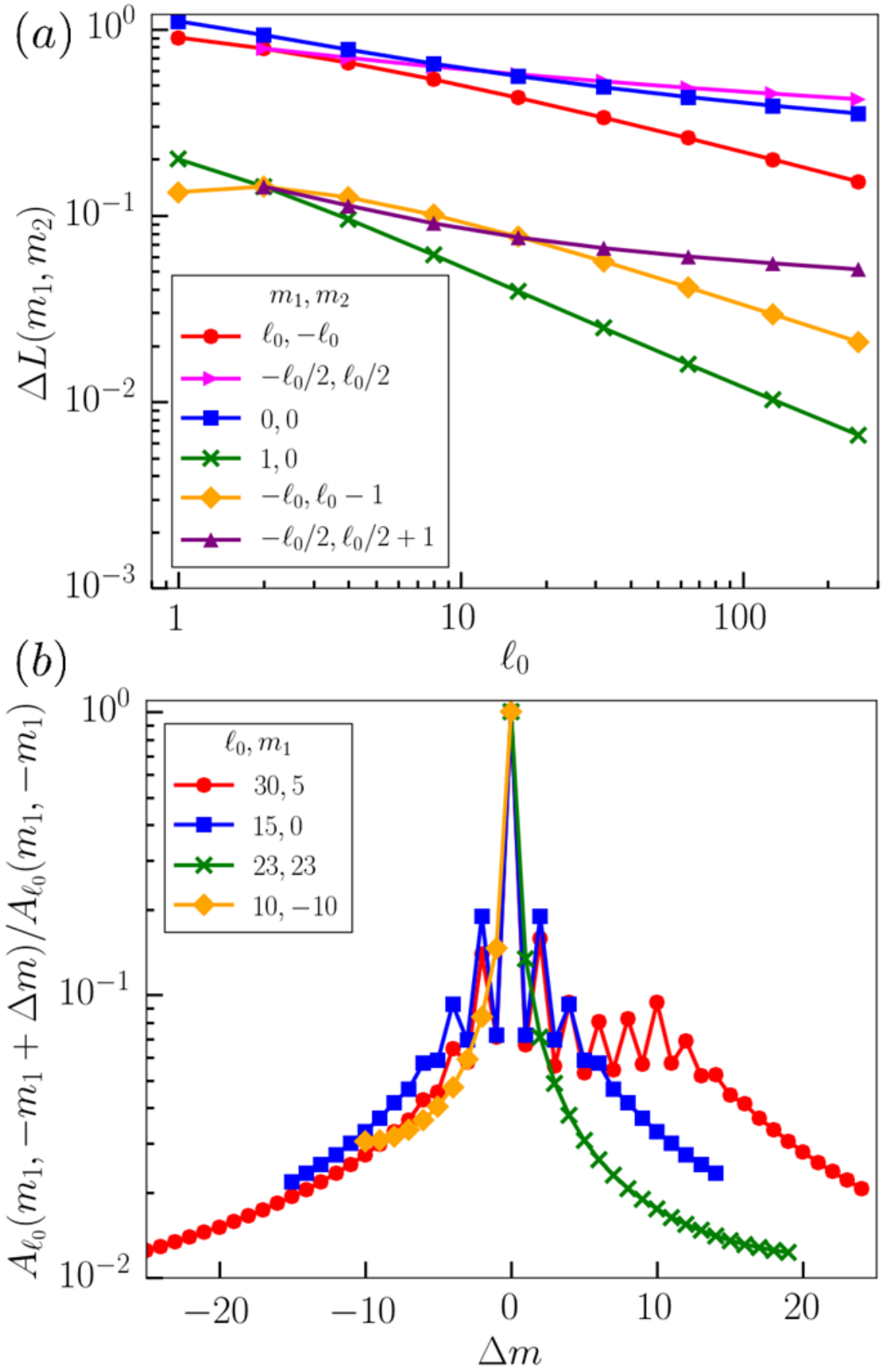}}
\caption{\label{fig:scattering} (a) The ratio $\Delta L$, calculated in Eq.~\ref{eq:loopratio}, of the two loop contributions to the correlation function in the ordered state (see  Eq.~\ref{eq:HFPropordered}) plotted for various values of $\ell_0$, $m_{1,2}$.  Note that this factor is small for all directions except when $m_1=-m_2$.   (b) The plotted ratio shows that the scattering function $A_{\ell_0}$ \ decays rapidly with increasing distance $|\Delta m|$ away from the special direction $m_1=-m_2$.  }
\end{figure} 

 We can check explicitly that Eq.~\ref{eq:newvertexloop2} does not contribute significantly.  The ratio of the two loop contributions for an arbitrary state $\bar{\Psi}$ with modes $\bar{\Psi}_m = \bar{a} c_m$ (see  Eq.~\ref{eq:orderedstatesol}) is given by 
\begin{align}
\Delta L & \equiv \left. \left| \parbox{8mm}{\begin{fmfgraph}(20,20) \fmfleft{l1,l2}
\fmfright{r1,r2} \fmf{plain,tension=3}{l1,v1,l2} \fmf{double,tension=0.4,right}{v1,v2,v1}\fmfdot{v1,v2} \fmf{plain,tension=3}{r1,v2,r2} \fmfv{decor.shape=circle,decor.filled=empty,decor.size=4}{l2,r2}
\end{fmfgraph}}      \right|  \right/ \left|  \parbox{5mm}{\begin{fmfgraph}(13,20) \fmfleft{i}
 \fmfright{o} \fmf{plain}{i,v,o} \fmf{double,tension=0.4}{v,v}\fmfdot{v}
\end{fmfgraph} \vspace*{-6mm}}  \right| \nonumber \\
& = \sum_{m_2} \frac{  48 \pi^2 A_{\ell_0}(m_1,m_2)}{(3-\delta_{m_2}) \alpha^{\ell_0}_{m_2,m_2}} \left[1+ \frac{2\pi R \sqrt{t_o}}{\sinh(2 \pi R \sqrt{t_o})} \right]\nonumber \\
& \approx  48 \pi^2 \sum_{m_2} \frac{  A_{\ell_0}(m_1,m_2)}{(3-\delta_{m_2}) \alpha^{\ell_0}_{m_2,m_2}} \equiv \sum_{m_2} \Delta L(m_1,m_2), \label{eq:loopratio}
\end{align}
where we have assumed $\pi R \sqrt{t_o} \gtrsim 1$ in the last line. In the finite-size limit $\pi R \sqrt{t_o} \ll 1$, the expression in the last line simply gets multiplied by a factor of 2.   We plot this ratio in Fig.~\ref{fig:scattering}(a) for a single non-zero ordered state mode $\bar{\Psi}_{m_2}= \bar{a} c_m$, for which $\Delta L = \Delta L(m_1,m_2)$.  To facilitate rapid computation of the Gaunt coefficients, we use a fast numerical algorithm \cite{numgaunt}. We find that the ratio is quite small ($\Delta L \ll 1$) for most values of $m_{1,2}$, except for values of $m_1$ that are close to $-m_2$.  This condition is the analog of the special directions discussed by Brazovskii \cite{Brazovskii}, where the external momenta of the loop contribution in Eq.~\ref{eq:newvertexloop} are aligned with the reciprocal lattice vectors of the patterned phase.  To check that $\Delta L$ indeed decreases rapidly away from these special directions, we plot in   Fig.~\ref{fig:scattering}(b) the ratio of scattering functions $A_{\ell_0}(m_1,-m_1+\Delta m)/A_{\ell_0}(m_1,-m_1)$, where $\Delta m$ is the distance away from the special direction.  We find that as $|\Delta m|$ increases, we get a rapid decay in the scattering function $A_{\ell_0}$.   When $\lambda_3 \neq 0$, the ordered state amplitudes in the loop correction might have a different scaling, as discussed previously.  The particular directions $c_m$ will also change.  However, since the summations over the internal propagators in the loops remain the same, we again expect to be able to neglect the loop in Eq.~\ref{eq:newvertexloop2}  relative to Eq.~\ref{eq:generalloop} even when $\lambda_3 \neq 0$, but a detailed check is beyond the scope of this paper.

Thus, we have (partially) justified our  neglect of the loop correction in Eq.~\ref{eq:newvertexloop} when computing the propagator in the ordered state.  This is also consistent with the Brazovskii analysis. So, going back to our equation for the propagator, we find
\begin{align}
(\parbox{6mm}{\begin{fmfgraph}(15,5) \fmfleft{v1} \fmfright{v2} \fmf{double}{v1,v2}
\end{fmfgraph}})^{-1} & \approx (\parbox{6mm}{\begin{fmfgraph}(15,5) \fmfleft{v1} \fmfright{v2} \fmf{vanilla}{v1,v2}
\end{fmfgraph}})^{-1}- \left[ \begin{fmfgraph}(15,10) \fmfleft{v1,v2} \fmfright{v3,v4} \fmf{vanilla}{v1,i1} \fmf{vanilla,tension=0}{i1,v2} \fmf{vanilla}{v3,i1} \fmf{vanilla,tension=0}{i1,v4} \fmfdot{i1} \fmfv{decor.shape=circle,decor.filled=empty,decor.size=4}{v2,v4}
\end{fmfgraph} +  \parbox{6mm}{\begin{fmfgraph}(15,27) \fmfleft{i}
\fmfright{o} \fmf{plain}{i,v,o} \fmf{double,tension=0.4}{v,v}\fmfdot{v}
\end{fmfgraph}}      \right] \nonumber \\
g_o^{-1}(\bm{\ell}_1,\bm{\ell}_2) & = M(\ell,\tau) (-1)^{m_1} \delta_{\ell_1-\ell_2} \delta_{m_1+m_2}  \nonumber \\
& \quad {} +\frac{ R^2}{2} \sum\limits_{\bm{\ell}_{3,4}}\gamma^{(4)}[\bar{\Psi}_{m_3}\bar{\Psi}_{m_4}+g_o(\bm{\ell}_3,\bm{\ell}_4)] , \label{eq:finalorderedprop}
\end{align} 
 where  we have used Eq.~\ref{eq:generalloop} for the loop correction.  It is clear that Eq.~\ref{eq:finalorderedprop}  reduces to  Eq.~\ref{eq:orderedprop} in the main text.   Finally, we may evaluate the inverse propagator at $\ell_1=\ell_2=\ell_0$ so that the inverse propagator just picks out the fluctuation-renormalized value of $\tau$, denoted by $t_o$: $g_o^{-1}=(-1)^{m_1}R^2 t_o(m_1) \delta_{m_1+m_2}$.  Note that $t_o$ will depend on the index $m_1$, due to the ordered state term in Eqs.~\ref{eq:treediagram2}. So,  Eq.~\ref{eq:finalorderedprop} reduces to:
\begin{align}
t_o(m_1) &  =  \tau+\frac{\lambda_4(-1)^{m_1} }{2}\sum\limits_{\ell,m,\bar{\ell}}  \Upsilon_{m_1,-m_1,0}^{\ell_0,\ell_0,\bar{\ell}} \Upsilon_{m,-m,0}^{\ell,\ell,\bar{\ell}}\langle \psi_{\ell}^{m}  \psi_{\ell}^{-m} \rangle  \nonumber \\ & \qquad \qquad  {}+ \frac{\lambda_4}{8 \pi} \sum_{m}\alpha_{m_1,m}^{\ell_0}|\bar{\Psi}_{m}|^2 \\
 &  =  \tau+\frac{\lambda_4 }{2}\sum\limits_{\ell,m,\bar{\ell}}  \frac{(-1)^{m+m_1}\Upsilon_{m_1,-m_1,0}^{\ell_0,\ell_0,\bar{\ell}} \Upsilon_{m,-m,0}^{\ell,\ell,\bar{\ell}}}{M(\ell,t_o(m))}  \nonumber \\ & \qquad \qquad  {}+ \frac{\lambda_4}{8 \pi} \sum_{m}\alpha_{m_1,m}^{\ell_0}|\bar{\Psi}_{m}|^2. \label{eq:generalt}
\end{align}
 After some rearrangement and relabelling of indices, we find the loop correction term that may be conveniently substituted into Eq.~\ref{eq:generalh}:
\begin{align}
  &\frac{\lambda_4}{2}\sum\limits_{\bm{\ell}_1,\bar{\ell}}   \Upsilon_{m_1,-m_1,0}^{\ell_1,\ell_1,\bar{\ell}} \Upsilon_{m,-m,0}^{\ell_0,\ell_0,\bar{\ell}}\langle \psi_{\ell_1}^{m_1}  \psi_{\ell_1}^{-m_1} \rangle=  \nonumber \\ & {}   {} \qquad \qquad(-1)^{m}   \left[t_o(m)-\tau-\frac{\lambda_4}{8 \pi}\sum_{m_1} \alpha^{\ell_0}_{m,m_1}|\bar{\Psi}_{m_1}|^2 \right]. \label{eq:corrfunterm}
\end{align}

 Now everything is in place to solve for the magnetic field modes $h_m$ just in terms of the ordered state modes $\bar{\Psi}_m$.  We substitute Eq.~\ref{eq:corrfunterm} into Eq.~\ref{eq:generalh} and find an equation for $h_m$ given just in terms of $\bar{\Psi}_m$ and $t_o$:
\begin{align}
h_m & =\frac{t_o(m) \bar{\Psi}^*_{m}}{4\pi} -  \frac{\lambda_4}{32 \pi^2}\sum_{m_1}\alpha^{\ell_0}_{m,m_1}|\bar{\Psi}_{m_1}|^2\bar{\Psi}^*_{m}  \nonumber \\
&  {} +\frac{\lambda_4}{24\pi} \sum_{m_{1,2,3},\bar{\bm{\ell}}} (-1)^{\bar{m}}\Upsilon^{\ell_0,\ell_0,\bar{\ell}}_{m_1,m_2 ,\bar{m}}   \Upsilon^{\ell_0,\ell_0,\bar{\ell}}_{m_3,m,-\bar{m}}   \prod_{i=1}^3  \bar{\Psi}_{m_i} \nonumber \\
& {} + \frac{\lambda_3}{8 \pi} \sum_{m_{1,2}} \Upsilon_{m,m_1,m_2}^{\ell_0,\ell_0,\ell_0} \bar{\Psi}_{m_1} \bar{\Psi}_{m_2} \\
&=\left[t   +\frac{\lambda_4(\delta_m-3)\alpha_{m,m}^{\ell_0}}{24\pi }|\bar{\Psi}_{m}|^2  \right] \frac{\bar{\Psi}_{m}^*}{4 \pi}+  \frac{\lambda_3}{4\pi} \sum_n(-1)^n  \nonumber \\
&\qquad  \times \Upsilon^{\ell_0,\ell_0,\ell_0}_{n,-n,0}\left[\bar{\Psi}_n\left(\frac{1}{2}-\delta_n\right)\delta_m+ \bar{\Psi}_0\delta_{m-n}\right]\bar{\Psi}_n^*, \label{eq:generalh2}
\end{align}
 In the second equality of Eq.~\ref{eq:generalh2}  we assumed the ordered state  modes $\bar{\Psi}_{m_i}$ cancel in pairs, so that $m_i = m$ for one of the three modes in the summations over $m_{1,2,3}$.  Note that this covers many possible cases because the sums are constrained so that $m_1+m_2+m_3=-m$.  Note that cubic term cannot be neglected in this equation.  It will influence the nature of the ordered states chosen by the system.

  It is clear from Eq.~\ref{eq:generalh2} that $\bar{\Psi}_m=0$ is a possible solution to the equation $h_m=0$.  However, there are also the non-trivial solutions with $\bar{\Psi}^*_m \neq 0$, corresponding to the patterned states.  These solutions have a simple form in the absence of a cubic term ($\lambda_3=0$ in Eq.~\ref{eq:generalh2}).  Dividing Eq.~\ref{eq:generalh2} by  $\bar{\Psi}^*_m$  yields a non-zero  solution to $h_m=0$:
\begin{equation}
\left|\bar{\Psi}_m \right|^2 = \bar{a}^2 |c_m|^2 = \frac{24 \pi t_o(m) }{\lambda_4(3-\delta_m) \alpha_{m,m}^{\ell_0}} . \label{eq:orderedstatesol}
\end{equation} 
The ordered state solutions in the presence of a cubic term are more complicated, but we may still choose the amplitude normalization $\bar{a}^2 = 4 \pi t_o/\lambda_4$ without loss of generality.  The ordered state amplitudes in Eq.~\ref{eq:orderedstatesol} depend on the function $t_o(m)$, which must be solved for using Eq.~\ref{eq:generalt}.
 This could be done numerically, but we will be interested in an analytically tractable approximation.  Hence, to make progress, we look for an isotropic approximation to Eq.~\ref{eq:generalt} and replace $t_o(m)$ with a constant $t_o$.  To do this, we must find some $m$-independent approximation to the coefficient $\alpha_{m_1,m}^{\ell_0}$ in Eq.~\ref{eq:generalt}.  The simplest solution is to average $\alpha_{m_1,m}^{\ell_0}$ over all external directions $m_1$: 
\begin{align}
\langle\alpha_{m_1,m}^{\ell_0} \rangle& =\frac{4\pi}{2\ell_0+1} \sum_{m_1,\bar{\ell}} (-1)^{m_1+m_2}\Upsilon^{\ell_0,\ell_0,\bar{\ell}}_{m_1,-m_1,0}\Upsilon^{\ell_0,\ell_0,\bar{\ell}}_{m_2,-m_2,0} \nonumber \\
& =\sqrt{4\pi}  (-1)^{m_2}\Upsilon^{\ell_0,\ell_0,0}_{m_2,-m_2,0} =  1. \label{eq:averagealpha}
\end{align}
It is also worth noting that the $\bar{\ell}=0$ term in the sum in the definition of $\alpha^{\ell_0}_{m_1,m}$ (Eq.~\ref{eq:alphadef}) contributes the most, as can be verified numerically.  Then, since $ 4\pi(-1)^{m_1+m_2}\Upsilon^{\ell_0,\ell_0,0}_{m_1,-m_1,0}\Upsilon^{\ell_0,\ell_0,0}_{m_2,-m_2,0}=1$ for any $m_{1,2}$,  replacing $\alpha_{m_1,m}^{\ell_0}$ with 1 in Eq.~\ref{eq:generalt} is a reasonable approximation.  After regularizing the propagator sum as in the disordered state calculation (Eq.~\ref{eq:L1final}), we find an $m$-independent solution for $t_o$:
\begin{equation}
t_o  =   \tau+   \frac{\lambda_4 \ell_0}{4R\sqrt{ t_o}} \, \coth(\pi R \sqrt{t_o})+ \frac{\lambda_4}{8 \pi} \sum_{m}|\bar{\Psi}_{m}|^2. \label{eq:finalt}
\end{equation}
 A similar neglect of the angular depedence of the mass term occurs in the Brazovskii analysis, where it has been shown that including the angular dependence does not substantially change the results \cite{anisotropicBrazo}.     Finally, note that the cubic term we have already thrown out (Eq.~\ref{eq:treediagramcubic}) vanishes in this approximation because it contributes the following to the renormalized parameter $t_o$:
\begin{equation}
\frac{ \lambda_3 }{2}\sum\limits_{m_{1}}  (-1)^{m_1}\Upsilon_{m_1,-m_1,0}^{\ell_1,\ell_1,\ell_0} \bar{\Psi}_{0} \propto  \delta_{\ell_0}, \label{eq:cubicvanishes}
\end{equation}
which vanishes for any $\ell_0>0$.  Similarly, the last loop contribution in Eq.~\ref{eq:HFPropordered} vanishes in this isotropic approximation.

 We now calculate the change in potential energy per unit area $\Delta \Phi$ in going from the disordered to the ordered state.  We recall that we ``turn on'' the ordered state by applying the field $h$, so that the ordered state modes $\bar{\Psi}_m=a c_m$  have their amplitudes $a$ increase from 0 to $\bar{a}$.  In a similar way, the renormalized parameter $t$ changes from $t_d$ to $t_o$. So, from Eq.~\ref{eq:generaldeltaphi} in the main text and Eqs.~\ref{eq:generalh2}, \ref{eq:orderedstatesol}, we find
\begin{align}
\Delta \Phi & =\sum_m \int_0^{\bar{a}} h_m \frac{\partial\bar{\Psi}_{m} }{\partial a}\mathrm{d}a \nonumber \\
& =\sum_m \int_0^{\bar{a}} \left[t a- \frac{\lambda_4(3-\delta_m)\alpha_{m,m}^{\ell_0}}{24 \pi} a^3 |c_m|^2 \right]\frac{ |c_m|^2}{4 \pi} \,\mathrm{d}a \nonumber \\
& +  \frac{\lambda_3}{4\pi} \sum_{m,n}\int_0^{\bar{a}}\Bigg\{(-1)^n \Upsilon^{\ell_0,\ell_0,\ell_0}_{n,-n,0}a^2  \nonumber \\
&\qquad  \qquad\times \left[c_n\left(\frac{1}{2}-\delta_n\right)\delta_m+ c_0\delta_{m-n}\right]c_n^*c_m \Bigg\} \mathrm{d}a\nonumber \\
& =\sum_m \frac{ |c_m|^2}{4 \pi} \left[\int_{t_d}^{t_o}ta  \frac{da}{dt} \,\mathrm{d}t   \right] +\Delta_4 +\frac{ \lambda_3 \Delta_3}{\sqrt{\lambda_4}} , \label{eq:DeltaPhi1}
\end{align} 
where we have changed variables from $a$ to $t$ in the left-over integral and found the quartic term contribution 
\begin{align}
\Delta_4 \equiv -\frac{   t_o^2}{24\lambda_4 }\sum_{m}  (3-\delta_m)\alpha_{m,m}^{\ell_0} |c_m|^4 
\end{align}
and a cubic term contribution
\begin{align}
\Delta_3 & \equiv\frac{ \sqrt{\pi}t_o^{3/2}c_0}{3\lambda_4}  \sum_{m}(3-2\delta_m)(-1)^m\Upsilon^{\ell_0,\ell_0,\ell_0}_{m,-m ,0}|c_m|^2.
\end{align}

  We now need the Jacobian factor $da/dt$. The two parameters $a$ and $t$ are connected via Eq.~\ref{eq:finalt}, generalized to the varying ordered state modes $\bar{\Psi}_m=a c_m$: \begin{align}
t& = \tau+\frac{ \lambda_4\ell_0}{4 R\sqrt{t}} \, \coth(\pi R  \sqrt{t})+ \frac{\lambda_4 a^2}{8 \pi} \sum_{m}  |c_m|^2, \label{eq:massgeneral}
\end{align}
which may be compared to Eq.~\ref{eq:massgeneralMT} in the main text. We now differentiate both sides of this equation with respect to $t$ and rearrange the terms to find our Jacobian $da/dt$:
\begin{align}
a \frac{da}{dt}& = \frac{4 \pi}{\lambda_4\sum_{m}  |c_m|^2}  \bigg\{1 + \nonumber \\ & \qquad \quad{} \frac{\lambda_4 \ell_0 \coth(\pi R \sqrt{t})}{8R t^{3/2}} \bigg[ 1+ \frac{2\pi R \sqrt{t}}{\sinh (2\pi R \sqrt{t})}\bigg]  \bigg\},
\end{align}
Substituting in the above expression  into Eq.~\ref{eq:DeltaPhi1} produces the final result:
\begin{align}
\Delta \Phi & = \Delta\Phi_0 +\Delta_4+\frac{ \lambda_3 \Delta_3}{\sqrt{\lambda_4}}  ,  \label{eq:finalPhi}
\end{align}
  where we have the contribution from the integral:\begin{align}
\Delta \Phi_0 & = \int_{t_d}^{t_o}\left[ \frac{t}{\lambda_4}+\frac{ \ell_0 \coth(\pi R \sqrt{t})}{8R t^{1/2}} \bigg[ 1+ \frac{2\pi R \sqrt{t}}{\sinh (2\pi R \sqrt{t})}\bigg]\right] \mathrm{d}t \nonumber \\ &= \frac{t_o^2-t_d^2}{2\lambda_4} + \frac{ \ell_0}{ 2\pi R^2}\Big[   \ln (\sinh(\pi R \sqrt{t})) \nonumber \\ &  \qquad \qquad \qquad  \qquad \quad  {}-\frac{\pi R \sqrt{t}}{2} \coth(\pi R \sqrt{t})  \Big]_{t=t_d}^{t_o}.
\end{align} 

In the planar limit, the free energy change in Eq.~\ref{eq:finalPhi} does not reduce to the Brazovskii result in an obvious way because it depends on the directions $c_m$ of the spherical harmonic modes.   However,  as in the planar case, we find that $\Delta \Phi$ becomes negative for a sufficiently negative parameter $\tau$.
 Equation~\ref{eq:finalPhi} may now be used in conjunction with the  solutions for the ordered states $\bar{\Psi}$ to find the most stable patterned phases.  
\end{fmffile}

\bibliographystyle{plain}
\bibliography{POP}

\end{document}